\documentclass[11pt]{article}

\PassOptionsToPackage{numbers,compress}{natbib}

\usepackage{microtype}
\usepackage{graphicx}
\usepackage{tabularx}
\usepackage{subcaption}
\usepackage{booktabs}
\usepackage[table]{xcolor}
\usepackage{microsoft-tech-report}
\usepackage{hyperref}
\usepackage{url}
\usepackage[utf8]{inputenc}
\usepackage{fix-cm}
\usepackage[T1]{fontenc}
\usepackage{amsmath}
\usepackage{amssymb}
\usepackage{amsfonts}
\usepackage{mathtools}
\usepackage{amsthm}
\usepackage{bm}
\usepackage{nicefrac}
\usepackage{latexsym}
\usepackage{enumitem}
\usepackage{multirow}
\usepackage{array}
\usepackage{longtable}
\usepackage{supertabular}
\usepackage{makecell}
\usepackage{multicol}
\usepackage[capitalize,noabbrev]{cleveref}
\usepackage{xspace}
\usepackage{pifont}
\usepackage{wrapfig}
\usepackage{float}
\usepackage{algorithm}
\usepackage{algpseudocode}
\usepackage{placeins}
\usepackage{lineno}

\usepackage{listings}
\usepackage{inconsolata}
\usepackage[most]{tcolorbox}

\lstdefinestyle{codeblock}{
    basicstyle=\ttfamily\small,
    frame=single,
    breaklines=true,
    columns=fullflexible,
    keepspaces=true,
    showstringspaces=false
}

\newtcblisting{promptbox}[1][]{
    enhanced,
    breakable,
    colback=blue!2!white,
    colframe=black!75,
    fonttitle=\bfseries\sffamily,
    coltitle=white,
    attach boxed title to top center={yshift=-2mm},
    boxed title style={
        colback=black!75,
        enhanced,
        arc=2pt
    },
    title={#1},
    listing only,
    listing options={
        basicstyle=\ttfamily\small,
        breaklines=true,
        breakatwhitespace=true,
        showstringspaces=false,
        columns=fullflexible,
        keepspaces=true,
    },
    boxrule=0.7pt,
    arc=3pt,
    left=4pt,
    right=4pt,
    top=6pt,
    bottom=4pt,
    drop shadow=black!10
}

\newcommand{\cnt}[1]{{\scriptsize (#1)}}

\definecolor{deltapos}{HTML}{0E8A4A}
\definecolor{deltaneg}{HTML}{B91C1C}

\techreportlabel{Microsoft Research}
\techreportshorttitle{RepoLaunch}

\hypersetup{
    colorlinks=true,
    linkcolor=msftblue,
    citecolor=msftblue,
    urlcolor=msftblue,
    pdftitle={
        RepoLaunch: Automating Build and Management of Code
        Repositories across Languages and Platforms
    }
}

\begin{document}
\thispagestyle{empty}


\noindent
\begin{minipage}[c]{0.5\linewidth}
\raggedright
\raisebox{-0.5\height}{\msftbrandmark}
\end{minipage}%
\begin{minipage}[c]{0.49\linewidth}
\raggedleft
{\msftdatefont\small\color{msftgray}October 2026}
\end{minipage}\par

\vspace{0.35em}
\noindent{\color{msftline}\rule{\linewidth}{0.8pt}\par}

\vspace{1.0em}

\begin{center}

{{\msfttitlefont\fontsize{21}{25}\selectfont\color{msftdark}
RepoLaunch: Automating Build and Management of Code Repositories\\
across Languages and Platforms$^{\ddagger}$\par}}

\vspace{1.25em}

{\normalsize\rmfamily\color{msftdark}
Kenan Li$^{\ast,\dagger}$ \hspace{0.6em}
Rongzhi Li$^{\ast}$ \hspace{0.6em}
Linghao Zhang$^{\ast}$ \hspace{0.6em}
Qirui Jin$^{\ast,\dagger}$ \hspace{0.6em}
Liao Zhu$^{\ast}$\\[-0.1em]
Xiaosong Huang \hspace{0.6em}
Geng Zhang \hspace{0.6em}
Yikai Zhang \hspace{0.6em}
Shilin He \hspace{0.6em}
Chengxing Xie\\[-0.1em]
Xin Zhang \hspace{0.6em}
Zijian Jin \hspace{0.6em}
Bowen Li \hspace{0.6em}
Chaoyun Zhang \hspace{0.6em}
Yu Kang \hspace{0.6em}
Yufan Huang\\[-0.1em]
Elsie Nallipogu \hspace{0.6em}
Saravan Rajmohan \hspace{0.6em}
Qingwei Lin \hspace{0.6em}
Dongmei Zhang\par
}

\vspace{0.22cm}

{\footnotesize\rmfamily\color{msftgray}
Microsoft\par
}

\end{center}

\vspace{0.45em}

\begin{msfttitlebox}
\setlength{\parindent}{0cm}
\setlength{\parskip}{0.14cm}
\raggedright
\nohyphens

\begin{abstract}
Language model (LM) agents have driven substantial progress in automated software engineering (SWE), yet building and testing software repositories at scale remains a largely manual and labor-intensive bottleneck.
In this work, we introduce \textbf{RepoLaunch}\textsuperscript{\textdaggerdbl}, a novel agentic framework that automatically resolves dependencies, compiles source code, and extracts test results across diverse programming languages and operating systems.
RepoLaunch achieves a \textbf{78\%} build success rate, outperforming the Python/Linux-only prior system by 18\%.
To demonstrate its application, we further present a fully automated pipeline for SWE dataset creation driven by RepoLaunch, which only requires human input at the task-design stage. 
RepoLaunch is open-sourced, and its automated task-generation pipeline has been adopted by several recent works on agentic benchmarking and training \citep{zhang2025swe, zeng2026glm, jin2026osbench}.
\end{abstract}

\vspace{0.14cm}

{\setlength{\parskip}{0.06cm}\small
{\msftmetalabel{Code}
\href{https://github.com/microsoft/RepoLaunch}
{github.com/microsoft/RepoLaunch}\par}}

{\setlength{\parskip}{0.06cm}\small
{\msftmetalabel{Dataset}
\href{https://huggingface.co/collections/SWE-bench-Live/swe-bench-live}
{huggingface.co/collections/SWE-bench-Live/swe-bench-live}\par}}

\vspace{0.08cm}

{\footnotesize\rmfamily\itshape\color{msftgray}

$^{\ast}$ RepoLaunch is a large-scale engineering project that the
first five authors have iteratively refined over a year. It is very
difficult to determine the order of their contributions. To the
extent possible, we believe that the first five authors should be
considered to have contributed equally.\par

$^{\dagger}$ Correspondence to Kenan Li
(\href{mailto:kenan.li.24@ucl.ac.uk}{kenan.li.24@ucl.ac.uk})
and Qirui Jin
(\href{mailto:qiruijin@gatech.edu}{qiruijin@gatech.edu}).\par

$^{\ddagger}$ This work was accepted as a poster paper at the
\textbf{NeurIPS 2026 Main Conference}.\par

}

\end{msfttitlebox}


\section{Introduction}

Building software repositories (repos) often requires substantial human effort. Repository configurations vary widely, documentation is frequently incomplete, and different platforms have specific dependency and compilation errors.
These challenges make repository setup a major barrier to scaling automated SWE workflows.

Recently, there has been growing interest in evaluating LM agents on real-world coding tasks within executable environments, as well as training LMs using execution feedback.
Notable benchmarks include the SWE-bench family~\citep{jimenez2023swe, yang2024multimodal, yang2025swesmith}.
Recent agentic training works on these SWE benchmarks include rejection-based supervised fine-tuning and reinforcement learning~\citep{zainullina2025guided, golubev2025training, cao2025skyrl}.
These advances underscore the need for scalable methods to build massive number of executable sandboxes and extract test results for evaluating agent performance, far exceeding manual capacity, motivating the need for agentic automation. 

GitTaskBench~\citep{ni2025gittaskbench} evaluates agents to solve GitHub issues on raw repos without a pre-built executable environment. It reports that in \textbf{65\%} of cases, agents fail to build executable environments even before starting to resolve issues.
The experiment reveals two key insights: (1) building repositories is itself a challenging problem that warrants dedicated study; (2) combining repository setup and issue resolution into a single task can obscure the distinct failure modes of each stage.
These observations further emphasize the importance of producing and managing pre-built executable repo sandboxes at scale.

Based on such demands, we propose RepoLaunch, an agent for fully automated repository build and test status management.
To the best of our knowledge, RepoLaunch is the \textbf{first} agent generalized to diverse programming languages and platforms, given only the language’s basic build/test prompts and a base OS image.
RepoLaunch autonomously explores a repository to install dependencies and compile the code.
However, building a repository alone is insufficient for practical SWE workflows.
Validating build results, managing repository state, tracking source code changes, and evaluating agent solutions all require reproducible re-build processes and structured extraction of test outcomes.
To address this gap, RepoLaunch formally defines and operationalizes the repository management task, including reproducible rebuild commands and standardized test status extraction.

Our experiments demonstrate that RepoLaunch can build and test repositories across nine programming languages on both Windows and Linux, including variants of Unix platforms such as Android and embedded systems, achieving an average build success rate of around 80\%.
This capability enables the automatic creation of large-scale, diverse SWE task instances that were previously infeasible to construct manually. 
We thus propose a RepoLaunch-automated SWE dataset creation pipeline with SWE-bench-Live / MultiLang and Windows task sets as demonstrations.
We further benchmark recent LMs and agentic systems on the multi-language, multi-OS tasks generated by RepoLaunch, offering insights into their current capabilities and limitations.

\begin{figure*}[t]
  \centering
  \includegraphics[width=\textwidth]{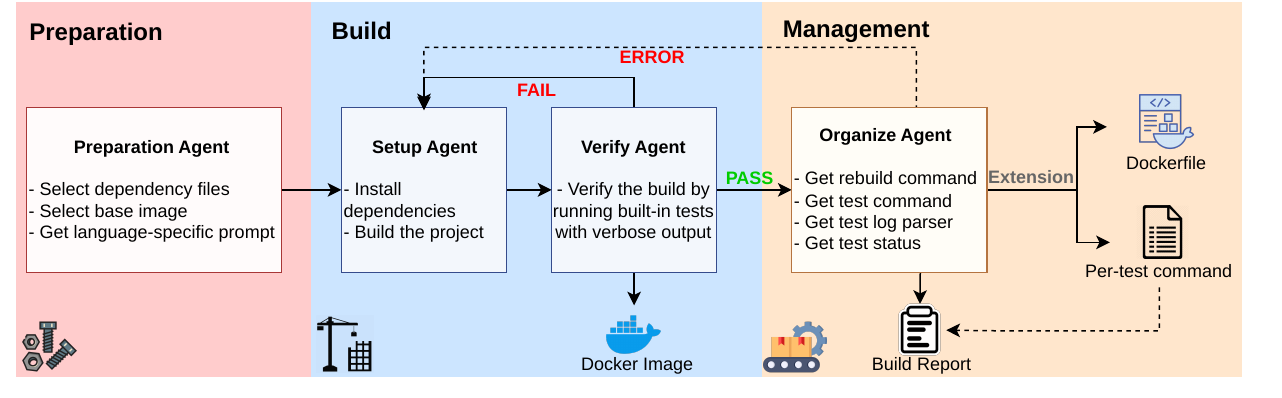}
  \caption{RepoLaunch Pipeline.}
  \label{fig:1}
\end{figure*}

The contributions of RepoLaunch are threefold:

\begin{enumerate}[nosep, leftmargin=1.5em]
\item \textbf{Formal definition of the repository build and management problem.}
Distinct from prior empirical studies on repository build, RepoLaunch formally defines the problem of \textit{repository build and test management}. It introduces an extensible framework with dedicated failure-pattern analysis to guide future improvements in agentic system design and LM training on this problem.

\item \textbf{The first solution generalized across arbitrary languages and platforms.}
Unlike prior empirical work that targets a specific language, a single platform, or only a subset of the problem for application-specific needs, RepoLaunch provides the first unified solution covering the full cycle of dependency resolution, build, and test-status management across arbitrary languages and platforms. We release RepoLaunch as an open-source tool so that users no longer need to reimplement repository build and test tooling for their individual applications.

\item \textbf{A fully automated pipeline for constructing SWE tasks for LM agent evaluation and training,} requiring human effort only at the task-design stage. As demonstrations, we release SWE-bench-Live/MultiLang and SWE-bench-Live/Windows, the \textbf{first} multi-language, multi-platform, and automatically-updating SWE dataset.

\end{enumerate}

\section{Related Works} 

\textbf{Rule-based automatic repo-build tools.} Oss-Fuzz-Gen \cite{ossfuzzgen} relies on predefined build instructions (such as ``./bootstrap.sh'', ``./configure'', ``make'') in specified files to build repos. Starter~\citep{starter} and Yeoman generator~\citep{yeomangenerator} fill Dockerfile templetes from project content. Pipreqs~\citep{pipreqs}, DockerizeMe~\citep{horton2019dockerizeme} and DockerGen~\citep{ye2021dockergen} infer dependencies from project content. SWE-MERA~\citep{adamenko2025swe} made a Python SWE benchmark with only hard-coded rules to build Python repos, with lower success rate in building repos and thus narrower task coverage compared to agentic methods below. These rule-based methods cannot cover the huge variants of project settings and structures especially for the noisy GitHub public repos, and are difficult for multi-language and multi-platform extension, illustrating the necessity of LM agents to inspect repos case by case.

\textbf{Initial efforts to build Python repos in Linux environment with agents.} Repo2Run \citep{hu2025llm} designed an agent for Python repos to install dependencies, change base images, execute tests and organize agents' commands into a Dockerfile as the result. SWE-rebench \citep{badertdinov2025swe} and OpenSWE \citep{fu2026davinci} designed agents to generate a Dockerfile with installation and test commands to create the executable environment of Python SWE tasks. However, setting up Python repos on Linux is very trivial, mostly with one simple command like "pip install ."; the test framework of Python is also unified and simple, with pytest covering most cases. Generalizing the automated build and test task to all languages and operating systems requires complex exploration process which such direct Dockerfile generation cannot cover. Therefore, RepoLaunch proposes a unified framework for cross-language and cross-platform adaptation, enables free agentic exploration inside the docker container, and designs a novel stage for test status management.
Notably, SWE-bench-Live \citep{zhang2025swe} and GLM-5 foundation model \citep{zeng2026glm} have used RepoLaunch to create SWE tasks for agentic evaluation and training.

\begin{figure*}[t]
  \centering
  \includegraphics[width=\textwidth]{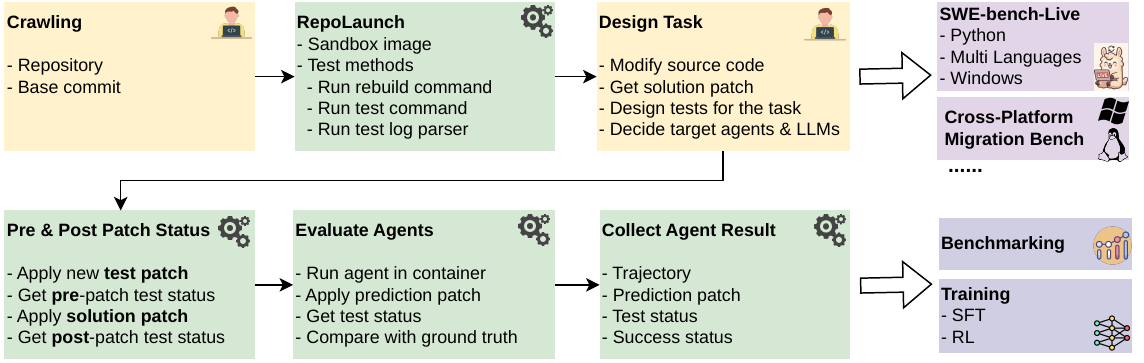}
  \caption{RepoLaunch-Automated SWE Dataset Creation. Green blocks are steps automated by RepoLaunch. In the top-right of the figure, \textit{Cross-Platform Migration Bench} (OS-bench~\citep{jin2026osbench}) is our latest work, also an SWE benchmark with task instances created by RepoLaunch.}
  \label{fig:2}
\end{figure*}

\section{RepoLaunch}

We address two core challenges in automating software repository management. The \textbf{first} is to \textbf{build} a repository and make it executable. However, a built repo is not yet usable and manageable without reproducible rebuild command and continuous test status monitoring. Therefore, the \textbf{second} problem of repo \textbf{management} is defined as producing the re-build command to install changed dependencies and compile the repo after source code edits, and generating the test command and parser to turn test output into structured test-status mapping. The solution to the second problem is critical for validating the build result, managing repo status and tracking source code changes. We formally define the two problems in Section \ref{problem_formulation}. To address them, we propose RepoLaunch -- a multi-agent workflow comprising three stages: \textbf{Preparation} (Section \ref{preparation_stage}), \textbf{Build}  (Section \ref{build_stage}) and \textbf{Management}  (Section \ref{management_stage}). We then demonstrate one application of RepoLaunch with an \textbf{automated pipeline for SWE dataset creation} (Section \ref{dataset_creation}).

\subsection{Problem Formulation}
\label{problem_formulation}

Let $R \in \mathcal{R}$ be a target repository.
Let $\mathcal{S}$ be the set of environment states and $\mathcal{C}$ the set of command sequences.
Define
\[
\delta : \mathcal{S} \times \mathcal{C} \to \mathcal{S}
\]
as the state--transition function, and
\[
\varepsilon : \mathcal{R} \times \mathcal{S} \to \{0,1\}
\]
as the verification function, where exit code $0$ denotes a successful build.
Let $\mathcal{B} \subseteq \mathcal{S}$ be the set of admissible base docker images, and let $S_{\emptyset} \in \mathcal{S}$ be the bare operating system.

In the first stage of \textbf{Preparation}, a base image $B \in \mathcal{B}$ is decided by executing a discovery sequence $C_B \in \mathcal{C}$:
\[
B = \delta(S_{\emptyset}, C_B), \qquad B \in \mathcal{B}.
\]

In the second \textbf{Build} stage, given $B$, a build command sequence $P \in \mathcal{C}$ is selected such that
\[
S_f = \delta(B, P)
\]
satisfies
\[
\varepsilon(R, S_f) = 0.
\]

In the third stage of \textbf{Management}, we fix the solution $(B^*, \mathcal{P}^*)$ from Build stage. A minimal rebuild subsequence needs to be extracted $\mathcal{P}_R \subseteq \mathcal{P}^*$ such that for any updated repository $R'$ yielding an updated base state $B'$,
\[
S_f' = \delta(B', \mathcal{P}_R)
\]
still satisfies
\[
\varepsilon(R', S_f') = 0.
\]

Let $C_{\mathrm{test}} \in \mathcal{C}$ be commands to print logs of per--test status.
Let $\mathcal{L}$ be the set of logs produced by
\[
\lambda : \mathcal{S} \times \mathcal{C} \to \mathcal{L},
\]
and the test log is thus defined as
\[
L_{\mathrm{test}} = \lambda(S_f, C_{\mathrm{test}}). 
\]

Let $\mathcal{T}$ be the set of test cases with status space
\[
\mathcal{O} = \{\mathrm{pass}, \mathrm{fail}, \mathrm{skip}\}.
\]
We also need to determine a parser
\[
\Pi : \mathcal{L} \to \mathcal{O}^{\mathcal{T}}
\]
mapping logs to outcome assignments.

\subsection{Preparation Stage}
\label{preparation_stage}

\textbf{Utilities}. RepoLaunch begins with two key utilities: (1) a \textbf{base image}, which provides a minimal operating system environment with basic dependencies for a given programming language, and (2) \textbf{language-specific instructions}, which offer build and test guidance for common frameworks of that language. See Appendix~\ref{base_image} for base image list and the instructions. These two utilities enable generalization of RepoLaunch across programming languages and platforms. Experiments show that the pre-built base images do not affect success rate of the agent, but only reduce trial steps and image size; however, the instructions do affect the success rate, as current LMs lack sufficient knowledge of the diverse frameworks for a language.

\textbf{Preparation Agent.} The Preparation Agent scans the repository’s file tree, read potential configuration and instruction files and retain relevant files for downstream stages. Based on the relevant files, the agent selects an appropriate base image, launches a \textbf{container} from the image, and copies the repository into it. The base image name, relevant file contents, and language-specific instructions are then passed as context to the next stages.

\subsection{Build Stage}
\label{build_stage}

\textbf{Setup Agent}. The Setup Agent uses the context from the Preparation Agent to install dependencies and compile the repository. It is equipped with a \textbf{Bash} tool for executing shell commands within the container and a \textbf{WebSearch} tool for retrieving external knowledge to resolve errors. After attempting to build the repository, the agent searches for the repository’s regression test commands. If a majority of test cases pass, the container is handed off to the \textbf{Verify Agent}.

\textbf{Verify Agent}. The Verify Agent validates the build outcome to mitigate LM hallucinations. It inspects the command history to identify test commands that report the status of \textbf{individual} test cases. If the majority of tests pass and tests related to core functionalities, especially integration tests pass, the container is committed as a new image. Otherwise, the agent reports the failure back to the Setup Agent for retry, until the maximum retrial limit, after which the build is marked as failed. 

Human verification of 100 instances that passed the Verify Agent shows only two cases had incomplete build, which were reliably caught and aborted by the subsequent agent.

\subsection{Management Stage}
\label{management_stage}

In the final stage, the \textbf{Organize Agent} extracts reusable artifacts from the previous records.

\textbf{Rebuild commands.} The Setup Agent’s command history may contain redundant or failed steps. The Organize Agent filters and optimizes this sequence to produce a minimal set of \textbf{rebuild commands} that can install the modified dependencies and recompile the project after code edits. These commands are then executed in the container to verify correctness. If issues are detected, the agent iteratively refines the commands until they execute successfully.

\textbf{Test commands and parser.} Similarly, the Organize Agent distills the Verify Agent’s command history into a minimal set of \textbf{test commands} that output the status of each test case. It also generates a \textbf{test log parser}—a Python function that parses the test output and returns a structured mapping from test case names to their statuses. When possible, the agent prefers test commands that produce structured output (e.g., JSON or XML) to reduce the difficulty of log parsing. Although RepoLaunch supports delegating failures in this stage back to the Setup Agent, we disabled this feedback loop in our experiments for efficiency.

\textbf{Per-test command (Optional).} In scenarios where fine-grained test control is needed downstream, the agent attempts to generate commands for executing individual test cases. It produces a Python script that returns a list of per-test commands from the test case list. The system samples and executes a subset of these commands, prompting the agent to verify and refine the script.

\textbf{Dockerfile (Optional).} If required, a Dockerfile can be reconstructed from the committed image’s layers. This step is deterministic and implemented via hard-coded logic.

\subsection{RepoLaunch-Automated SWE Dataset Creation}

\label{dataset_creation}

This subsection highlights RepoLaunch’s broader application beyond repository setup. As discussed earlier, coding LMs require large-scale, diverse datasets of software engineering (SWE) tasks for benchmarking and training. Traditionally, creating such datasets—particularly those involving repository builds and test status extraction—has required extensive manual engineering effort. RepoLaunch automates this process end-to-end.

As illustrated in Figure~\ref{fig:2}, for new SWE task creation,  researchers only need to consider task design. Specifically, they must: (1) select repositories at relevant commits to clone, (2) design new tasks by modifying the repository (e.g., introducing bugs or incomplete features), and (3) introduce new tests to validate whether a coding agent has resolved the task. RepoLaunch builds the repository, saved as an image, and extracts rebuild\&test commands and a test log parser for structured test results.

To evaluate a task instance, RepoLaunch applies the test patch and the ground-truth solution patch separately. It executes the rebuild and test commands and parses the resulting logs to obtain the test statuses before and after the ground-truth solution patch. This yields the status of both existing regression tests and newly added task-specific tests. An LM coding agent can attempt the task in the executable environment from the saved image and submit its solution patch. RepoLaunch then compares the agent’s test outcomes with those of the ground-truth solution to determine whether the agent has successfully solved the task.

In addition to enabling automated benchmarking, the agent’s full execution trajectory can be saved for further analysis or agent training. This pipeline allows human effort to focus only on the creative task design, while the repetitive and time-consuming steps of repo build and test management are all delegated to RepoLaunch.

\section{Experiment}

\subsection{Evaluation of RepoLauch}

The evaluation of RepoLaunch includes three experiments:

\textbf{Exp 1. RepoLauch-automated SWE dataset creation.} To demonstrate the application of RepoLauch-automated SWE dataset creation, RepoLaunch is used to generate multi-language and multi-platform SWE task instances contributed to SWE-bench-Live \citep{zhang2025swe}, which only had Python tasks on Linux before, from real GitHub issues and pull requests. We measure success rates in the Build and Management stages and analyze failure patterns from the real distribution of GitHub repositories and commits. 

\textbf{Exp 2. Comparison with related works on a human-labeled benchmark.} RepoLaunch is compared against several related works to build sandbox images and extract test statuses for existing human-curated SWE benchmarks whose 'FAIL\_TO\_PASS' labels were manually verified. We use task instances from SWE-bench-Verified \citep{jimenez2023swe}, SWE-bench-Multilingual \citep{yang2025swesmith} and SWE-bench-Pro \citep{deng2025swepro}. For each repository, we select three task instances with commits farthest apart in time to evaluate reproducibility and robustness to dependency drift. The final benchmark contains 152 instances from 8 languages (see Appendix \ref{comparison_instance_list} for full task list).

We then compare RepoLaunch against several baselines on this benchmark: SWE-agent \citep{yang2024swe}, a general-purpose SWE agent with only bash and file-editing tools, repo2run \citep{hu2025llm}, the first repo-build agent prior to our work, limited to Python base images and hard-coded commands (e.g., pip install, pytest), CXXCrafter\citep{yu2025cxxcrafter}, a similar C/C++-only repo build agent, pipreqs\citep{pipreqs}, a rule-based Python build method, railpack\citep{railpack}, the latest rule-based repo build method equipped with human-written rules for several languages. To the best of our knowledge, there is no other agent-based method so far capable of handling multiple languages/platforms other than RepoLaunch.

\textbf{Exp 3. Ablation study.} We further use this benchmark to isolate the contributions of certain modules of RepoLaunch -- language-specific image, prompt and the search, file-editor tools, and to evaluate the influence of different LM providers on RepoLaunch's performance.

\textbf{Hyper Parameters.} The Setup Agent is allowed up to 60 steps on Linux and 90 on Windows due to increased complexity. The Verify Agent is capped at 20 steps. Rebuild, test, and parser generation stages are each limited to 20 steps. Each bash command has a 30-minute timeout. For comparison, the maximum step of SWE-agent and repo2run is set to 120 due to context and rate limits. If not specified, the default LM to drive the agents is \textbf{\textit{GPT-5-Thinking-Medium}}.

\textbf{Computing resources.} Experiments were done with 3 Linux machines (96 cores, 1TB RAM each) and 3 Windows machines (16 cores, 16GB RAM each) in 30 days.

\subsection{SWE Task Creation from GitHub}

RepoLaunch proposes the \textbf{first} automated method to create SWE tasks across \textbf{multiple} programming languages and platforms. The task creation pipeline is as follows:

\begin{enumerate}[nosep]
\item \textbf{Crawl:} Collect GitHub issues with associated pull requests (PRs).

\item \textbf{Verification:} Use an LM to filter out unreliable task descriptions.

\item \textbf{Split:} Use an LM to classify issues as general or Windows-specific.

\item \textbf{Launch:} Run RepoLaunch to build and test the repository at the commit preceding the pull request.

\item \textbf{Validation:} Retain instances with test cases that fail (or skip / do not run) before the PR and pass after.

\end{enumerate}

Refer to Appendix \ref{pipeline} for details of the pipeline.

\textbf{SWE-bench-Live on Multiple Languages} (SWE-bench-Live/MultiLang). It includes tasks in C/C++, C\#, Java, Golang, JavaScript/TypeScript (JS/TS) and Rust. Due to resource and time constraints, we sampled pull requests in July-August 2025 and March-April, July-August 2026, yielding 1077 task instances from 431 repositories --- surpassing the original SWE-bench-Multilingual benchmark (300 instances from 41 repos), whose smaller dataset size shows the difficulty of manually building multi-language repos other than Pythons.

\textbf{SWE-bench-Live for Windows-Specific Tasks} (SWE-bench-Live/Windows). It includes tasks in Python, C/C++, C\#, Java, Go, JS/TS, and Rust. Prior benchmarks focused only on Linux environments, lacking evaluation on Windows-specific issues and Windows command line operations. Recent leakage and accidental deletion of user data by agents such as OpenClaw \citep{openclaw} further highlight the significance of Windows-specific ability. We crawled and sampled 600 Windows specific issues from January 2025 to May 2026.

\textbf{Evaluation of LM agents on SWE-bench-Live / MultiLang \& Windows}. To assess the utility of RepoLaunch datasets, we also evaluate some popular SWE agents (the open-source SWE-agent \citep{yang2024swe}, OpenHands \citep{wang2024openhands} and close-source ClaudeCode \citep{claudecode}) and latest LMs (GPT-5.5-Medium, GPT-5.2-Medium, Claude-4.5-Sonnet, Gemini-3-flash, Deepseek-v4-Pro, Deepseek-v3.1-terminus) on the tasks created, delivering new insights to the community on the latest progress of SWE agents and LMs. Agentless \citep{xia2024agentless} was not evaluated as it only supports Python; Kimi \citep{team2025kimi} was not evaluated due to its extremely low rate limit. Since the three agents do not support Windows containers, we implemented a minimal Windows-compatible agent equipped with the same toolset as SWE-agent and OpenHands, named as Win-agent for Windows benchmarking. The max step of rollout for these agents is set at 100.

\section{Result}
\label{Result}

\begin{table*}
  \centering
  \begin{tabular}{llllll}
    \toprule
    \textbf{Language}  & \textbf{Total} & \textbf{Verification} & \textbf{Build} & \textbf{Management}  & \textbf{Validation}  \\
    \midrule
    Python* & 1200 & 693 / 57.8\% & 528 / 76.2\% & -- & 100 / 18.9\%  \\
    JS/TS & 1300 & 493 / 37.9\% & 439 / 89.0\% & 405 / 92.3\% & 219 / 54.1\%  \\
    C/C++ & 2200 & 997 / 45.3\% & 822 / 82.4\% & 710 / 86.4\% & 253 / 35.6\%  \\
    C\# & 1400 & 437 / 30.5\% & 402 / 92.0\% & 383 / 95.3\% & 136 / 35.5\%  \\
    Java & 1200 & 508 / 42.3\% & 396 / 78.0\% & 372 / 93.9\% & 160 / 43.0\%  \\
    Go & 700 & 300 / 42.9\% & 217 / 72.3\% & 206 / 94.9\% & 138 / 67.0\%  \\
    Rust & 1300 & 585 / 45.0\% & 422 / 72.1\% & 342 / 81.0\% & 171 / 50.0\%  \\
    Windows & 600 & 322 / 53.7\% & 227 / 70.5\% & 182 / 80.1\% & 065 / 35.7\%  \\
    \bottomrule
  \end{tabular}
  \caption{\label{citation-guide}
    The number and ratio of instances retained at each stage from the last stage for each language on Linux and the overall result on Windows. Python* data were referred from an old version of RepoLaunch in prior works contributed to the SWE-bench-Live Python subset in July and August 2025, where test log parsing adopted the rule-based pytest-format parser of SWE-bench \citep{jimenez2023swe}, so it does not have the Management step.
  }
  \label{performance}
\end{table*}

\subsection{Task Creation of SWE-bench-Live / MultiLang \& Windows}
\label{task_creation_result}

\textbf{Proportion of instances retained at each stage of SWE task creation} is shown in Table \ref{performance}, including the success rate of RepoLaunch in the Build and Management stages, and the retain rate in the verification and validation stages. Results on Linux and Windows images are basically the same, though the build success rate on Windows is a bit lower, showing the difficulty of SWE on Windows platform.

\textbf{Token usage and cost of LM API}. Average token usage and cost for the Preparation, Build, and Management stages by language are shown in Table \ref{cost_token}. Launching one instance costs about \$1.7 on average. C/C++ is the most expensive, reflecting its greater difficulty as an older language, with diverse frameworks and complex dependency, compilation, and testing commands. In contrast, repositories that can be resolved with a single command, such as "make" or "go test -v", cost much less, around \$0.15. While these seemingly simple cases might appear suitable for hard-coded rules, our trajectory analysis shows that about one-third of them still encounter build or test errors. Since it is difficult to distinguish truly simple cases from deceptive ones in advance, we did not use rule-based methods and instead adopted a unified agentic approach to handle each repo case by case.

\textbf{Scalability.} The success rate and cost as repo size grows are shown in Figure \ref{scalability}. RepoLaunch maintains consistent performance as repo sizes increase. The higher success rate for small repos is mainly attributable to their generally simple settings. The higher success rate for larger repos, on the other hand, is mainly because they are well documented and well maintained, with no apparent bugs. The low cost at both ends aligns with the high success rate, indicating the build/test easiness.

\textbf{Failure pattern analysis of RepoLaunch.} To guide future improvement on SWE agents for automatic repo build\&management, we conduct comprehensive failure pattern analysis of RepoLaunch. We manually labeled the execution trajectories of RepoLaunch for 100 failed instances in the Build stage and 100 in the Management stage to get the failure categories, described in Appendix~\ref{failure_pattern_desc}.

\begin{figure*}[t]
  \centering
  \includegraphics[width=0.96\linewidth]{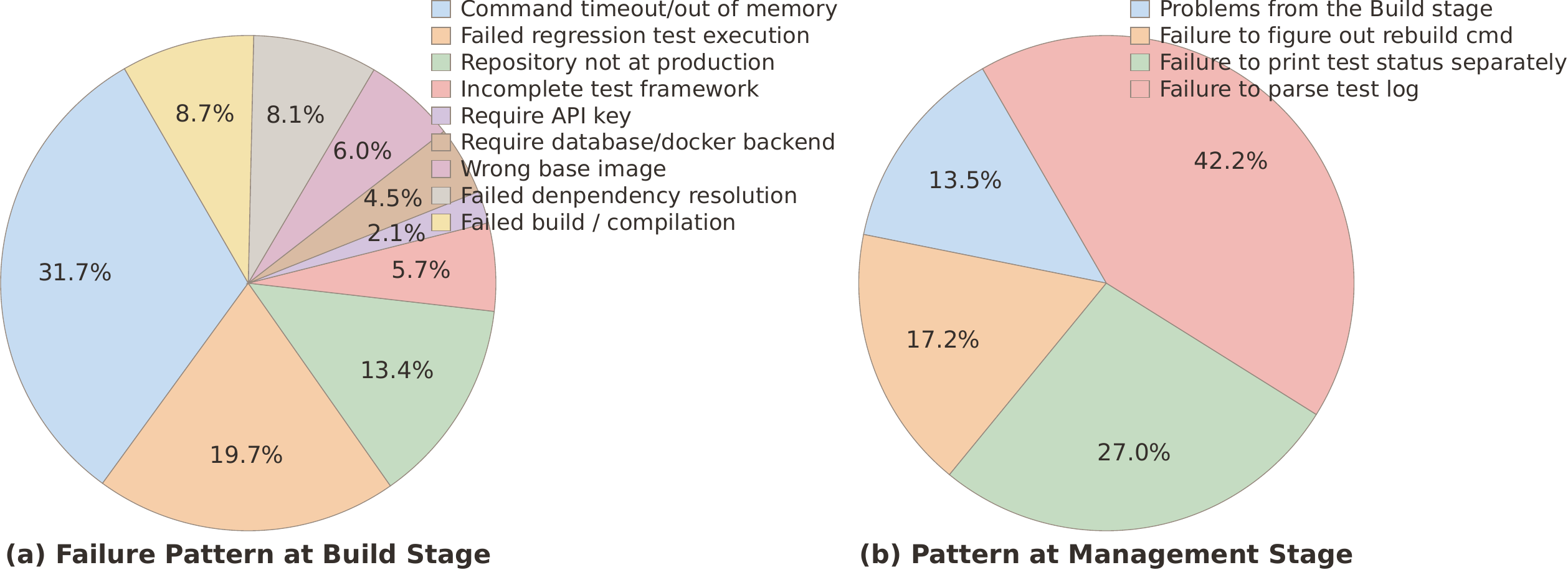}
  \caption {The proportion of different failure patterns in the Build and the Management stage .}
  \label{failue_pattern}
\end{figure*}

Then we use a Reasoning LM (GPT-5.2-Thinking-Medium) to label the left instances with these categories. The discrepancy between the proportions labeled by the LM and those by human does not exceed 20\%. Final proportions of each failure pattern on the 6 languages evaluated on Linux are shown in Figure \ref{failue_pattern}. For the per-language and Windows result please refer to Table \ref{perlang_pattern}. The failure pattern on Windows also aligns with that on Linux -- command timeout / out of memory (OOM) and failure to run regression test are the major factors. On Windows the difficulty to resolve dependencies is also a major challenge. The timeout / OOM failure is because the large repos in our test set are intrinsically resource-intensive. Due to large-scale parallelism requirements and limited resources, the timeout limit is set only 30 min. Given more time and resources, such failure can mostly be resolved.

\textbf{Cross-platform transferability.} The cross-platform transferability of RepoLaunch's design is further tested on Android with 92\% success and embedded systems with 88\% success. For experimental details please refer to Appendix \ref{transferability}.

\subsection{Evaluation Result of LMs and Agents on SWE-bench-Live/MultiLang\&Windows}

Table \ref{tab:benchmark-comparison} compares the average success rate of each LM and each agent, as well as the best LM-agent combination for each language. Table \ref{eval_res} shows the success rate of each LM-agent combination on each language on the MultiLang benchmark. The benchmarking results reveal:

1) The new generation of LMs becomes stronger with the highest success rate over 35\%, compared to prior results of around 15\% of older LMs on SWE-bench-Live Python tasks \citep{zhang2025swe}.

2) The performance gap between LMs is much greater than that between agents. 

3) No LM or agent can master all the languages, but each can excel in at least one language. Java appears to be the easiest, whilst Rust is the most difficult. The success rate on a language may depend on the amount of LM training data available for that language.

The benchmarking result of SWE-bench-Live/Windows is shown in Table \ref{windows_benchmarking_table}. Compared to Linux result, the success rates on Windows drop for all LMs. Claude remains robust on Windows command line environment and Windows-specific code implementation. 

Further failure pattern analysis is conducted on these coding agents' trajectories on Linux and Windows. As this is not the main topic of this paper, we put it in Appendix \ref{pattern_solution}.

\subsection{Comparison with Related Works}

To the best of our knowledge, we found no open-source works that include the second Management stage for test case extraction. Therefore, we  were only able to compare RepoLaunch with the general-purpose SWE-agent on both the Build and Management stages, while for other repo build methods, we only compared their Build stage performance.

\textbf{Comparison with SWE-agent.} For evaluation of SWE-agent, its initial prompt is set as "build a repo in a container, and write a bash script that rebuilds the repo and extracts test statuses". The evaluation of SWE-agent includes two criteria -- \textbf{Build} and \textbf{Management\&Validation} (\textbf{Manage\&Val}). Successful \textbf{Build} requires regression tests can run, while successful \textbf{Management\&Validation} requires successful extraction of "FAIL\_TO\_PASS" test cases from a SWE-bench instance. The successful extraction of "FAIL\_TO\_PASS" test cases requires 1) the repo is fully not partially built; 2) successful rebuild and 3) accurate test status extraction, which is a reliable criterion to ensure both build and management are entirely correct. The performance comparison of RepoLaunch with SWE-agent is shown in Table~\ref{performance_comp_table}. The test cases extracted by the two agents are manually aligned with the ground truth test names to eliminate formatting differences of the agent-written test log parsers. RepoLaunch surpasses the baseline SWE-agent on \textit{all} languages and on \textit{both} the \textbf{Build} and \textbf{Manage\&Val} stages, with overall \textbf{Build} success rate of 79.6\% and \textbf{Manage\&Val} success rate of 81.8\%. 

Notably, SWE-bench instances were manually validated to ensure that each includes non-empty FAIL\_TO\_PASS test cases. Therefore, RepoLaunch’s high validation retention rate of 91.5\% on SWE-bench instances suggests that the lower retention rate observed when constructing SWE-bench-Live instances from native GitHub PRs, as shown in Table \ref{citation-guide}, is largely because many of these PRs genuinely lack FAIL\_TO\_PASS test cases. 

\begin{table*}[t]
\centering

\begin{minipage}[c]{0.30\textwidth}
\centering

\subcaptionbox{}{\begin{tabular}{ll}
\hline
\textbf{Agent} & \textbf{Success} \\
\hline
SWE-agent & \textbf{29.6\%} \\
ClaudeCode & 29.2\% \\
OpenHands & 27.9\% \\
\hline
\end{tabular}}

\vspace{1.0em}

\subcaptionbox{}{\begin{tabular}{ll}
\hline
\textbf{LLM} & \textbf{Success} \\
\hline
GPT-5.5 & \textbf{35.8\%} \\
Deepseek-v4 & 30.0\% \\
Claude-4.5 & 28.5\% \\
GPT-5.2 & 25.1\% \\
Gemini-3 & 24.2\% \\
Deepseek-3.1 & 20.6\% \\
\hline
\end{tabular}}

\end{minipage}
\hfill
\begin{minipage}[c]{0.66\textwidth}
\centering

\subcaptionbox{}{\begin{tabularx}{\linewidth}{
    l>{\raggedright\arraybackslash}Xl}
\hline
\textbf{Split} & \textbf{Best combination} & \textbf{Success} \\
\hline
Java & GPT-5.5 + SWE-agent & \textbf{50.0\%} \\
JS/TS & GPT-5.5 + SWE-agent & 47.5\% \\
C\# & GPT-5.5/Deepseek-v4 + SWE-agent & 44.9\% \\
Go & Claude-4.5 + ClaudeCode & 43.8\% \\
C/C++ & Claude-4.5 + OpenHands/SWE-agent & 43.8\% \\
Rust & Gemini-3 + SWE-agent & 37.5\% \\
Windows & Claude-4.5 + Win-Agent & 30.0\% \\
\hline
\end{tabularx}}

\end{minipage}

\caption{Comparison of benchmarking results.
(a) Success rate of agents averaged across the 6 LMs and 6 languages;
(b) Success rate of LMs averaged across the 4 agents and 7 splits
(6 splits of 6 languages on Linux and 1 Windows task split);
(c) For each split, the best-performing combination of agents and LMs.}
\label{tab:benchmark-comparison}

\end{table*}

\textbf{Performance comparison with other open-source repo build methods} is shown in Table~\ref{performance_comp_table_build}. repo2run's original criterion to judge the build as successful is that all the test functions can be \textit{imported}, which is too lax. For fair comparison, here we consider the build of repo2run as successful when \textit{more than half of the tests pass}. RepoLaunch still achieves state-of-the-art performance on all languages whilst ensuring generalizability, outperforming previous python-linux-only method repo2run by 18\%. Generally speaking, agentic methods surpass rule-based methods, mostly because agentic methods can refine their solutions based on the observed build/test errors while rule-based methods cannot. 

\textbf{Cost comparison.} Table~\ref{cost_comp_table} compares RepoLaunch with several repo build\&test methods on average LM API cost per instance to build Python repos. As these methods use LMs to different extents in the test Management stage, where some use mixed rule-based methods, the LM cost at the Management stage cannot be compared directly. As shown in the table, RepoLaunch incurs relatively low API costs while delivering state-of-the-art performance.

\subsection{Ablation Study}

\textbf{Contribution of individual modules.} We ablate each of the four modules in RepoLaunch to assess its individual contribution: the language-specific prompt, the language-specific image, the web-search tool, and the string-replace file editor tool. Linux ablation is conducted on the same 152-instance benchmark in Appendix~\ref{comparison_instance_list}, while Windows ablation randomly samples 200 instances from the 500 Windows instances in Section~\ref{task_creation_result}. Table~\ref{ablation} shows the four modules all have significant contributions and are more useful on Windows, because Windows commands are more complicated with additional settings and formatting requirements, and repos usually require additional Windows-specific dependencies. Language-specific prompt helps more on the Management stage as it instructs on how to direct test output to structured output file (JSON or XML) to make test log parsing easier. Language-specific image significantly reduces average steps and increases success rate in the Build stage as it makes dependency installation much easier, but it does not contribute to subsequent stages. As task creation in SWE-bench-Live must preserve repository bugs, we did not include the file-editing tool by default in our previous experiments.

\textbf{Influence of LM backbone.} We further evaluate the robustness of RepoLaunch when using different LM providers in Table~\ref{language_model_robustness} on the benchmark in Appendix~\ref{comparison_instance_list}. RepoLaunch still outperforms other agents when driven by different LM providers. When using Claude-Sonnet-5 which is a specialized model for SWE problems, RepoLaunch achieves 90.8\% \textbf{Build} success and 92.0\% \textbf{Manage\&Val} success. These results highlight the value of stronger agentic tools and SWE-focused model training. Due to budget constraints, we were unable to use Claude models for large-scale evaluation.

\section{Conclusion and Future Directions}

This work introduces RepoLaunch, the first agent that automatically builds and manages code repositories across programming languages and platforms, surpassing prior repo build\&test baselines. We use RepoLaunch to create SWE-bench-Live / MultiLang and Windows datasets to demonstrate its application in automating SWE dataset creation for agentic benchmarking and training.  

RepoLaunch offers dedicated failure pattern analysis and has exposed interfaces for additional agentic tool integration and agent trajectory collection, guiding future improvements on this problem such as experience learning -- integrating database retrieval tool to build agent memory and training LM to learn from launch results with rejection-based fine-tuning / reinforcement learning.

RepoLaunch has been released as an open-source tool for the community. To date, it has received 2k+ Git clones per month and has led to 30+ meetings with researchers and engineers from the academia and industry.

\section*{Limitations}

RepoLaunch is a multi-turn free agent, which costs a lot of tokens. Future works might focus on reducing LM calls such as using hard-coded test log parser to cover common test frameworks first. 

The rule-based plus LM-based screening and filtering of GitHub tasks may contain bias. However, this is the common practice of automated SWE task creation in SWE-rebench \citep{badertdinov2025swe}, SWE-MERA \citep{adamenko2025swe} and OpenSWE \citep{fu2026davinci}. To date, no better solution exists. Nevertheless, SWE-bench-Live \citep{zhang2025swe} shows that LM-based verification has high precision (92\%) compared to manual labeling on SWE-bench-Verified \citep{jimenez2023swe}. We would keep the discussion open in the community for better solutions.

Running the large repositories from GitHub requires intensive computing. Due to computing resource limit, only 1077 + 65 instances have been created for SWE-bench-Live so far. But this is not the end. We will continue expanding the task set over time, incorporating the latest SWE problems from GitHub to support and benefit the community in agentic benchmarking and training.

\section*{Code Availability}
RepoLaunch: \url{https://github.com/microsoft/RepoLaunch};

SWE-bench-Live: \url{https://github.com/microsoft/SWE-bench-Live};

Win-agent: \url{https://github.com/njukenanli/Win-Agent}.

\section*{Data Availability}
SWE-bench-Live dataset: \url{https://huggingface.co/collections/SWE-bench-Live/swe-bench-live};

The sandbox docker images for each task instance are hosted on Dockerhub with namespace \textit{starryzhang};

SWE-bench-Live leaderboard: \url{https://swe-bench-live.github.io/}.

\section*{Acknowledgments}

RepoLaunch is a large-scale engineering project that the first five authors, Kenan Li, Rongzhi Li, Linghao Zhang, Qirui Jin and Liao Zhu, have iteratively refined over a year. It is very difficult to determine the order of their contributions. To the extent possible, we believe that the first five authors should be considered to have contributed equally. We also thank our supervisors, Chaoyun Zhang, Xin Zhang, and Yu Kang for their valuable guidance and assistance in identifying resources.


\bibliographystyle{unsrtnat}
\bibliography{custom}


\appendix

\appendix

\section{RepoLaunch Utilities}

\label{base_image}

\subsection{Base Image List}

If official base images for a language on an OS is provided, we directly refer to it; otherwise we built it from scratch. The Docker images below whose names start with "custom" are base images that we built manually.

\begin{longtable}{l|l|l}
    \toprule
    \textbf{Language} & \textbf{OS} & \textbf{Docker Images / Versions} \\
    \midrule
    \endfirsthead

    \multicolumn{3}{c}%
    {{\bfseries \tablename\ \thetable{} -- continued from previous page}} \\
    \toprule
    \textbf{Language} & \textbf{OS} & \textbf{Docker Images / Versions} \\
    \midrule
    \endhead

    \midrule
    \multicolumn{3}{r}{{Continued on next page}} \\
    \endfoot

    \bottomrule
    \endlastfoot

    Python & Linux & python:3.\{6,7,8\dots14\} \\
           & Windows & python:3.\{9,10,11\}-windowsservercore-ltsc2022 \\
           &         & python:3.\{12,13,14\}-windowsservercore-ltsc2025 \\
    \midrule
    C/C++  & Linux & mcr.microsoft.com/devcontainers/cpp:1-ubuntu-\{20,22,24\}.04 \\
           & Windows & custom/windows\_server:ltsc\{2019,2022\}\_cmake\_ninja\_only \\
           &         & custom/windows\_server:ltsc2022\_cmake\_ninja\_vsbuildtools\_cl\_msbuild \\
    \midrule
    C\#    & Linux & mcr.microsoft.com/dotnet/sdk:\{6,7,8,9,10\}.0 \\
           & Windows & mcr.microsoft.com/dotnet/sdk:\{8,9\}.0-windowsservercore-ltsc2019 \\
           &         & mcr.microsoft.com/dotnet/sdk:\{8,9,10\}.0-windowsservercore-ltsc2022 \\
    \midrule
    Java   & Linux & eclipse-temurin:\{11,17,21\}-jdk-noble \\
           & Windows & eclipse-temurin:\{11,17,21\}-jdk-windowsservercore-ltsc2022 \\
    \midrule
    JS/TS  & Linux & node:\{18,20,22,24,25\} \\
           & Windows & custom/windows\_server:ltsc2025\_nvm \\
    \midrule
    Go     & Linux & golang:1.\{19,20\dots25\} \\
           & Windows & golang:1.\{19,20\dots25\}.0-windowsservercore \\
    \midrule
    Rust   & Linux & rust:1.\{70,71\dots90\} \\
           & Windows & custom/rust-windows:1.\{70,75,80,85,90\} \\
\end{longtable}

\subsection{Language-Specific Instructions}

As the Instructions are very long, we only provide an example of C/C++ Windows instructions here. For full Instructions please refer to source code.

\begin{promptbox}[C/C++ Windows Build Instructions]
This is a windows server image with git, choco, cmake, ninja, and vsbuildtools2022 with cl.exe and msbuild installed.

Test these packages with: git --version; choco --version; cmake --version; ninja --version; cl.exe; msbuild --version;
The VSbuildtools2022 has already been installed by: choco install visualstudio2022buildtools --package-parameters "--add Microsoft.VisualStudio.Workload.NativeDesktop --add Microsoft.VisualStudio.Component. VC.Tools.x86.x64 --add Microsoft.VisualStudio.Component. Windows11SDK.22621 --add Microsoft.VisualStudio.Component. VC.CMake.Project --includeRecommended";

You need to figure out how to install other dependencies yourself. You can use web search if you are not sure.
choco is the preferred way to install packages. For example, choco install -y mingw, choco install -y llvm
Some dependencies may not be supported by choco and have to be installed through its official source. For example, git clone https://github.com/microsoft/vcpkg.git; ./vcpkg/bootstrap-vcpkg.bat; 
$env:VCPKG_ROOT = "C:\path\to\vcpkg"; $env:PATH = "$env:VCPKG_ROOT;$env:PATH";

Some dependency install examples using choco:

-- Qt with MSVC: 
(installing MSVC has been done)
choco install -y aqt --no-progress; refreshenv; 
aqt install-qt --outputdir C:\Qt windows desktop 6.8.0 win64_msvc2019_64;
aqt install-tool --outputdir C:\Qt windows desktop tools_qtcreator;
aqt install-tool --outputdir C:\Qt windows desktop tools_cmake;

-- Qt with MinGW: 
choco install -y qt6-base-dev mingw; refreshenv;

The installed packages often do not know the existence of each other. You need to link them manually if errors occur.

Examples to build a repo:

# Configure with CMake:

cmake -S . -B build -G Ninja -DCMAKE_BUILD_TYPE=Release -DCMAKE_CXX_STANDARD=23`

Use 20/17/11 if your project requires it

# Force a compiler if needed:

 -  GCC: `-DCMAKE_CXX_COMPILER=g++`
 -  Clang: -DCMAKE_CXX_COMPILER=clang++`

# Build the project:
  - `cmake --build build --parallel`

# Run tests:
  - `ctest --test-dir build --output-on-failure`

# Run the app:
  - `./build/<target_name>`
\end{promptbox}

\begin{promptbox}[C/C++ Windows test instructions]
- For **GoogleTest (gtest)**, run:
  ./your_test _binary --gtest_output= json:reports/gtest-results.json

- For **Catch2**, run:
  ./your_test_binary --reporter json > reports/catch2-results.json

- For **doctest**, run:
  ./your_test_binary --reporters=json > reports/doctest-results.json

- For **CppUTest**, run:
  ./your_test_binary -oj > reports/cpputest-results.json

- For **CTest (CMake test runner)**, run:
  ctest --output-log reports/ctest-results.json --output-junit reports/ctest-junit.xml
  (CTest does not produce JSON directly, but you can use the JUnit XML output)

- For **Boost.Test**, run:
  ./your_test_binary --report_level=detailed --log_format=JSON --log_sink=reports/boost-results.json
\end{promptbox}

\section{SWE-bench-Live / MultiLang \& Windows Task Creation Pipeline}
\label{pipeline}

\begin{enumerate}[nosep]
\item \textbf{Crawl.} Filter GitHub public repos with stars > 1k. Then, filter these repos with the proportion of any of the six languages > 60\%. Crawl merged pull requests (PRs) that are related to some issues during a specific time period.

\item \textbf{Verification.} Use a reasoning LM (GPT-5-Thinking-Medium here) to verify whether the test patch contains requirements that are not specified in the issue description, or the issue descritpion directly contains ground truths either in natural language or code. 

\item \textbf{Split.} Use a Reasoning LM (GPT-5-Thinking-Medium here) to categorize issue descriptions as either general/Linux or Windows-specific. Windows-specific problems should be built inside Windows images, instead of the Linux images adopted by previous SWE benchmarks.

\item \textbf{Launch.} Use RepoLaunch to build the sandbox images, extract the rebuild \& test commands and the test log parser for each task instance before the PR was merged.

\item \textbf{Validation.} Apply the test patch of the PR and run rebuild commands, test commands and test log parser to get the test status \textbf{before} the solution patch. Apply the test \textbf{and} solution patch of the PR and run again to get the test status \textbf{after} the solution patch. Compare the test status before and after the solution patch. Extract \textbf{PASS\_TO\_PASS} test cases that pass both before and after the solution patch. Extract \textbf{FAIL\_TO\_PASS} test cases that fail (or skip, or are missing) before but pass after.
For languages that require compilation to execute, if code snippets are not implemented correctly or contain errors, related tests will fail to build, or the entire repository may become uncompilable due to failing static code checks.
In such case the related test cases will only run and pass after the related code snippet is implemented correctly.
As such test cases can still test whether a code snippet is implemented correctly, we still take them as "FAIL\_TO\_PASS".
For evaluation of coding agents, their submitted prediction patch must pass both \textbf{PASS\_TO\_PASS} and \textbf{FAIL\_TO\_PASS} to be considered correct.

\end{enumerate}

\section{Token Usage and Cost of LM API}
Table \ref{cost_token} tracks the average token usage and cost per instance for each language and for the Preparation, Build and Management stages respectively. Python data were referred from an old version of RepoLaunch in prior works contributed to SWE-bench-Live Python tasks, which used a rule-based pytest-format parser to parse test logs, so it does not have a Management stage.

\begin{table*}
  \centering
  \begin{tabular}{llllllll}
  \hline
  & \textbf{C/C++} & \textbf{C\#} & \textbf{Java} & \textbf{JS/TS} & \textbf{Go} & \textbf{Rust} & \textbf{Python} \\
  \hline
  \rowcolor{gray!20}
  \textbf{Preparation Stage}&&&&&&&\\
  Input tokens & 75.4k & 70.8k & 43.2k & 95.7k & 87.2k & \textbf{116.1k} & 74.3k \\
  Output tokens & 412 & 476 & 264 & 405 & 406 & \textbf{580} & 430 \\
  USD cost & 0.073 & 0.075 & 0.043 & 0.073 & 0.075 & \textbf{0.103} & 0.071 \\
  \rowcolor{gray!20}
  \textbf{Build Stage}&&&&&&&\\
  Input tokens & \textbf{2619.8k} & 1090.0k & 561.0k & 1065.6k & 620.3k & 856.7k & 854.3k \\
  Output tokens & \textbf{7893} & 2762 & 2239 & 1860 & 1595 & 2341 & 2413 \\
  USD cost & \textbf{1.823} & 0.717 & 0.446 & 0.643 & 0.449 & 0.579 & 0.592 \\
  \rowcolor{gray!20}
  \textbf{Management Stage}&&&&&&&\\
  Input tokens & 743.6k & 952.0k & 643.8k & 904.4k & \textbf{1070.1k} & 765.9k & 0 \\
  Output tokens & 4839 & 6285 & 5043 & \textbf{6712} & 4410 & 5093 & 0 \\
  USD cost & 0.788 & 0.991 & 0.745 & \textbf{1.014} & 0.864 & 0.801 & 0 \\
  \rowcolor{gray!20}
  \textbf{Sum of 3 Stages}&&&&&&&\\
  Input tokens & \textbf{3438.8k} & 2112.8k & 1248.0k & 2065.7k & 1777.6k & 1738.7k & 928.6k \\
  Output tokens & \textbf{13144} & 9523 & 7546 & 9877 & 6411 & 8014 & 2843 \\
  USD cost & \textbf{2.684} & 1.783 & 1.234 & 1.730 & 1.388 & 1.483 & 0.663 \\
  \hline
  \end{tabular}
  \caption{\label{cost_token}
    Average token usage and cost of RepoLaunch per instance for each language and each stage. The experiments were done using GPT-5-Thinking-Medium. Python data were referred from an old version of RepoLaunch in prior works contributed to SWE-bench-Live Python tasks, which adopted a rule-based pytest-format parser to parse test logs, so it does not involve the Management stage .}
\end{table*}

\section{Scalability}

\begin{figure*}[t]
  \centering
  \includegraphics[width=\textwidth]{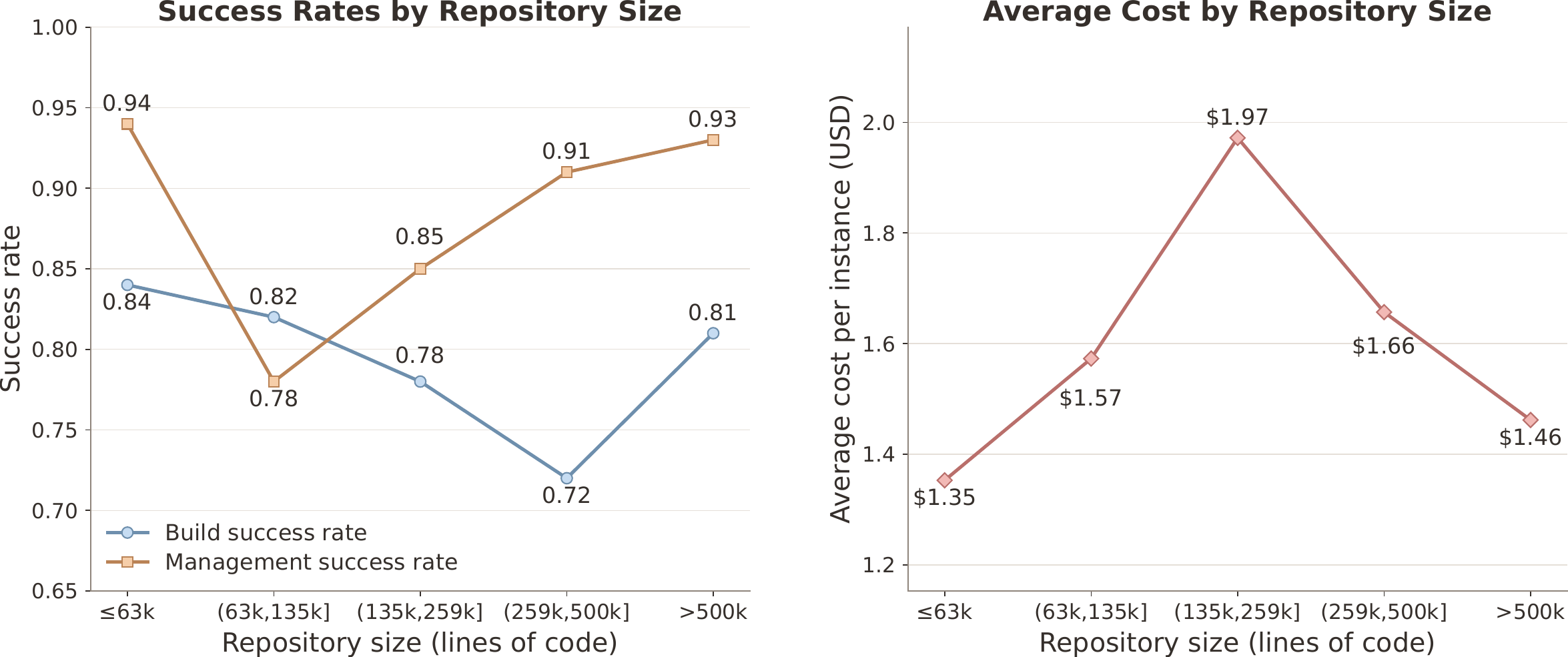}
  \caption{Scalability of RepoLaunch when repo size grows. Each of the five repo size ranges contains 75 repos.}
  \label{scalability}
\end{figure*}

Figure \ref{scalability} shows the relationships between success rate and repo size (line of codes) and between cost per instance and repo size (line of codes) on the Linux and Windows instances run in \textbf{Section} \ref{task_creation_result}.

\section{Failure Pattern of RepoLaunch}
\subsection{Failure Pattern Description}

\label{failure_pattern_desc}

\textbf{The Build stage.}

\begin{enumerate}[nosep]
\item \textbf{Repository not at production}. The repository at the current commit is not ready for production. The source code has some problems so it is impossible to build the project if we don't modify the source code.

\item \textbf{Incomplete test framework.} The repository does not contain regression tests, or does not have a complete test framework that can run regression tests, or cannot output the test status of each test case one by one, so the LM agent cannot verify whether the build is successful or not.

\item \textbf{Require API key.} The repository requires an API key to run tests.

\item \textbf{Require database/docker backend.} The repository requires a database/docker server installed in the container environment to run. This is the most difficult case when the environment is a single container.

\item \textbf{Command timeout / out of memory.} The build or test command of the repository timed out or cannot be done under the current memory constraints, so we cannot verify whether the repository can be built and tested. This is probably due to the repository being too large.

\item \textbf{Wrong base image.} The LM coding agent selected the wrong base image.

\item \textbf{Failed denpendency resolution.} The LM coding agent does not know how to install some of the dependencies.

\item \textbf{Failed build / compilation.} The LM coding agent does not know how to build / compile the repository.

\item \textbf{Failed regression test execution.} The repository has complete test framework that can print the status of each test case, but the LM coding agent failed to figure out how to run regression tests in the repository.
\end{enumerate}

\textbf{The Management stage.}

\begin{enumerate}[nosep]
\item \textbf{Problems from the Build stage.} All the problems introduced or brought from the Build stage (all the failure patterns in the Build stage), which failed to be rejected by the Build stage. Note this category includes but \textit{is not limited to} the sole build failure.

\item \textbf{Failure to figure out rebuild cmd.} The agent fails to figure out how to re-build the repository in the limited 30 steps.

\item \textbf{Failure to print test status separately.} The repository has complete test framework that can print the status of each test case, but the agent fails to figure out how to run tests and print the status of EACH test case one by one in the limited 30 steps.

\item \textbf{Failure to parse test log.} The agent fails to write the correct test log parser to extract the status of each test case in the limited 30 steps.
\end{enumerate}

\subsection{Failure Pattern of RepoLaunch per Language and for Windows}

\label{failure_pattern_per_lang}
Please refer to Table \ref{perlang_pattern}.

\begin{table*}
  \centering
  \begin{tabular}{llllllll}
    \hline
    \textbf{Failure pattern}  & \textbf{C/C++} & \textbf{C\#} & \textbf{Java}  & \textbf{TS/JS} & \textbf{Go} & \textbf{Rust} & \textbf{Win}   \\
    \hline
    \rowcolor{gray!20}
    \textbf{Build Stage} &  &  &  &  &  &  &  \\
    Repo not at production & \textbf{9.7\%} & 12.4\% & 14.5\% & \textbf{20.7\%} & 10.8\% & \textbf{17.6\%} & 11.3\%  \\
    Incomplete test framework & 7.8\% & 6.2\% & 4.8\% & 4.6\% & 4.3\% & 9.9\% & 1.4\% \\
    Require API key & 0.0\% & 0.0\% & 0.0\% & 4.6\% & 3.6\% & 0.0\% & 0.7\%  \\
    Require database / docker backend & 0.0\% & 9.9\% & 1.2\% & 5.1\% & 8.6\% & 3.3\% & 1.4\%  \\
    Command timeout / out of memory & \textbf{61.2\%} & \textbf{25.9\%} & 10.8\% & 20.3\% & \textbf{36.7\%} & \textbf{22.0\%} & \textbf{20.4\%}  \\
    Wrong base image & 3.9\% & 8.6\% & 13.3\% & 2.3\% & 5.0\% & 12.1\% & 12.7\%  \\
    Failed dependency resolution & 5.8\% & 19.8\% & \textbf{16.9\%} & 3.7\% & 10.1\% & 3.3\% & \textbf{20.2\%}  \\
    Failed build / compilation & 4.8\% & 0.0\% & \textbf{21.7\%} & 5.1\% & 3.6\% & 16.5\% & 12.0\%  \\
    Failed regression test execution & 6.8\% & \textbf{17.3\%} & \textbf{16.9\%} & \textbf{33.2\%} & \textbf{17.3\%} & 15.4\% & \textbf{19.7\%}  \\
    \rowcolor{gray!20}
    \textbf{Management Stage} &  &  &  &  &  &  &  \\
    Problems from the Build stage & 16.7\% & 7.9\% & 10.9\% & 6.6\% & 17.2\% & 20.9\% & 11.5\% \\
    Fail to figure out rebuild cmd & 19.4\% & 7.9\% & 15.6\% & 13.1\% & 10.3\% & 25.6\% & 30.8\% \\
    Fail to print test status separately & 22.2\% & 20.6\% & 20.3\% & 39.3\% & 27.6\% & 16.3\% & 13.5\% \\
    Fail to parse test log & \textbf{41.7\%} & \textbf{63.5\%} & \textbf{56.3\%} & \textbf{41.0\%} & \textbf{44.8\%} & \textbf{37.2\%} & \textbf{44.2\%} \\
    \hline
  \end{tabular}
  \caption{\label{perlang_pattern}
    Failure Pattern per Language and for Windows in Build and Management stages. The highest two values in each column for Build stage and the highest one value in each column for Management stage are bolded.
  }
\end{table*}

\section{Comparison of Methods on Repo Build and Test Management}

\subsection{The Benchmark and Task instance\_ids for Method Comparison}

\label{comparison_instance_list}

We compare different methods -- agentic or rule-based -- to build repositories at three commits farthest apart in time for each repo and extract "FAIL\_TO\_PASS" test cases at the specified commits. The repo-commit pairs come from SWE-bench-Verified, SWE-bench-Multilingual and SWE-bench-Pro multilingual instances, whose "FAIL\_TO\_PASS" test cases were human-verified to ensure a fair comparison. The benchmark we build contains 152 instances from 8 languages. The instance\_ids of the instances for comparison are as follows:
\\

\setlength{\tabcolsep}{0.1mm}
\small
\begin{longtable}{p{5cm}p{3cm}p{5cm}p{1.5cm}}
\toprule
\textbf{instance\_id} & \textbf{language} & \textbf{instance\_id} & \textbf{language} \\ 
\midrule
\endfirsthead

\multicolumn{4}{c}%
{{continued from previous page}} \\
\toprule
\textbf{instance\_id} & \textbf{language} & \textbf{instance\_id} & \textbf{language} \\
\midrule
\endhead

\midrule \multicolumn{4}{r}{{Continued on next page}} \\ \midrule
\endfoot

\bottomrule
\endlastfoot
jqlang\_\_jq-2235 & C & apache\_\_druid-13704 & Java \\
jqlang\_\_jq-2681 & C & apache\_\_druid-14136 & Java \\
jqlang\_\_jq-2919 & C & apache\_\_druid-16875 & Java \\
micropython\_\_micropython-10095 & C & apache\_\_lucene-11760 & Java \\
micropython\_\_micropython-13039 & C & apache\_\_lucene-12626 & Java \\
micropython\_\_micropython-15898 & C & apache\_\_lucene-13704 & Java \\
redis\_\_redis-9733 & C & google\_\_gson-1014 & Java \\
redis\_\_redis-11631 & C & google\_\_gson-2061 & Java \\
redis\_\_redis-13338 & C & google\_\_gson-2479 & Java \\
valkey-io\_\_valkey-790 & C & javaparser\_\_javaparser-4538 & Java \\
valkey-io\_\_valkey-1499 & C & javaparser\_\_javaparser-4561 & Java \\
valkey-io\_\_valkey-1842 & C & projectlombok\_\_lombok-2792 & Java \\
fmtlib\_\_fmt-1683 & C++ & projectlombok\_\_lombok-3371 & Java \\
fmtlib\_\_fmt-3248 & C++ & projectlombok\_\_lombok-3697 & Java \\
fmtlib\_\_fmt-3901 & C++ & reactivex\_\_rxjava-7597 & Java \\
nlohmann\_\_json-4237 & C++ & axios\_\_axios-4731 & JS/TS \\
caddyserver\_\_caddy-4774 & Go & axios\_\_axios-5316 & JS/TS \\
caddyserver\_\_caddy-6051 & Go & axios\_\_axios-6539 & JS/TS \\
caddyserver\_\_caddy-6411 & Go & babel\_\_babel-13928 & JS/TS \\
gin-gonic\_\_gin-1805 & Go & babel\_\_babel-15445 & JS/TS \\
gin-gonic\_\_gin-3227 & Go & babel\_\_babel-16130 & JS/TS \\
gin-gonic\_\_gin-4003 & Go & facebook\_\_docusaurus-8927 & JS/TS \\
gohugoio\_\_hugo-12171 & Go & facebook\_\_docusaurus-9897 & JS/TS \\
gohugoio\_\_hugo-12448 & Go & facebook\_\_docusaurus-10309 & JS/TS \\
gohugoio\_\_hugo-12768 & Go & immutable-js\_\_immutable-js-2005 & JS/TS \\
hashicorp\_\_terraform-34580 & Go & immutable-js\_\_immutable-js-2006 & JS/TS \\
hashicorp\_\_terraform-34900 & Go & mrdoob\_\_three.js-25687 & JS/TS \\
hashicorp\_\_terraform-35611 & Go & mrdoob\_\_three.js-26589 & JS/TS \\
prometheus\_\_prometheus-9248 & Go & mrdoob\_\_three.js-27395 & JS/TS \\
prometheus\_\_prometheus-12874 & Go & preactjs\_\_preact-2757 & JS/TS \\
prometheus\_\_prometheus-15142 & Go & preactjs\_\_preact-3567 & JS/TS \\
future-architect\_\_vuls-9aa0d8 & Go & preactjs\_\_preact-4436 & JS/TS \\
future-architect\_\_vuls-ca3f6b & Go & vuejs\_\_core-11589 & JS/TS \\
future-architect\_\_vuls-bff6b7 & Go & vuejs\_\_core-11870 & JS/TS \\
gravitational\_\_teleport-0cb341 & Go & vuejs\_\_core-11915 & JS/TS \\
gravitational\_\_teleport-b4e7cd & Go & protonmail\_\_webclients-e65cc5 & JS/TS \\
gravitational\_\_teleport-eefac6 & Go & protonmail\_\_webclients-4feccb & JS/TS \\
navidrome\_\_navidrome-874b17 & Go & protonmail\_\_webclients-ac23d1 & JS/TS \\
navidrome\_\_navidrome-6b3b4d & Go & NodeBB\_\_NodeBB-0e07f3 & JS/TS \\
navidrome\_\_navidrome-09ae41 & Go & NodeBB\_\_NodeBB-1ea948 & JS/TS \\
flipt-io\_\_flipt-406f93 & Go & NodeBB\_\_NodeBB-a5afad & JS/TS \\
flipt-io\_\_flipt-7161f7 & Go & element-hq\_\_element-web-33e8ed & JS/TS \\
flipt-io\_\_flipt-86906c & Go & element-hq\_\_element-web-53a9b6 & JS/TS \\
astropy\_\_astropy-7166 & Python & element-hq\_\_element-web-107772 & JS/TS \\
astropy\_\_astropy-13579 & Python & tutao\_\_tutanota-518182 & JS/TS \\
astropy\_\_astropy-14995 & Python & tutao\_\_tutanota-f3ffe17 & JS/TS \\
django\_\_django-7530 & Python & tutao\_\_tutanota-f373ac & JS/TS \\
django\_\_django-13933 & Python & burntsushi\_\_ripgrep-2209 & Rust \\
django\_\_django-17087 & Python & burntsushi\_\_ripgrep-2576 & Rust \\
matplotlib\_\_matplotlib-13989 & Python & nushell\_\_nushell-12901 & Rust \\
matplotlib\_\_matplotlib-24570 & Python & nushell\_\_nushell-13246 & Rust \\
matplotlib\_\_matplotlib-26466 & Python & nushell\_\_nushell-13831 & Rust \\
mwaskom\_\_seaborn-3069 & Python & sharkdp\_\_bat-562 & Rust \\
mwaskom\_\_seaborn-3187 & Python & sharkdp\_\_bat-2393 & Rust \\
pallets\_\_flask-5014 & Python & sharkdp\_\_bat-3108 & Rust \\
psf\_\_requests-1142 & Python & tokio-rs\_\_axum-682 & Rust \\
psf\_\_requests-2317 & Python & tokio-rs\_\_axum-1119 & Rust \\
psf\_\_requests-6028 & Python & tokio-rs\_\_axum-2096 & Rust \\
pydata\_\_xarray-2905 & Python & tokio-rs\_\_tokio-4384 & Rust \\
pydata\_\_xarray-4695 & Python & tokio-rs\_\_tokio-6603 & Rust \\
pydata\_\_xarray-7393 & Python & tokio-rs\_\_tokio-7139 & Rust \\
pylint-dev\_\_pylint-4551 & Python & uutils\_\_coreutils-6377 & Rust \\
pylint-dev\_\_pylint-6528 & Python & uutils\_\_coreutils-6682 & Rust \\
pylint-dev\_\_pylint-8898 & Python & uutils\_\_coreutils-6731 & Rust \\
pytest-dev\_\_pytest-5262 & Python & astral-sh\_\_ruff-15309 & Rust \\
pytest-dev\_\_pytest-7324 & Python & astral-sh\_\_ruff-15394 & Rust \\
pytest-dev\_\_pytest-10356 & Python & astral-sh\_\_ruff-15626 & Rust \\
scikit-learn\_\_scikit-learn-9288 & Python & sympy\_\_sympy-11618 & Python \\
scikit-learn\_\_scikit-learn-14053 & Python & sympy\_\_sympy-17655 & Python \\
scikit-learn\_\_scikit-learn-26323 & Python & sympy\_\_sympy-24661 & Python \\
sphinx-doc\_\_sphinx-7440 & Python & qutebrowser\_\_qutebrowser-77c355 & Python \\
sphinx-doc\_\_sphinx-8721 & Python & qutebrowser\_\_qutebrowser-9b71c1 & Python \\
sphinx-doc\_\_sphinx-11510 & Python & qutebrowser\_\_qutebrowser-e15d26 & Python \\
internetarchive\_\_openlibrary-fad4a4 & Python & ansible\_\_ansible-189fcb & Python \\
internetarchive\_\_openlibrary-308a35 & Python & ansible\_\_ansible-de01db & Python \\
internetarchive\_\_openlibrary-8a5a63 & Python & ansible\_\_ansible-1c06c4 & Python \\
\label{comparison}

\end{longtable}

\subsection{Evaluation Results of Different Methods on Repo Build and Management}

Table \ref{performance_comp_table} and Table \ref{performance_comp_table_build} compare the performance of different methods in building and testing SWE-bench repos, while Table~\ref{cost_comp_table} compares the LM API cost of several relevant agentic methods in building repos. RepoLaunch achieves state-of-the-art performance on both repo build and test management with a low cost.

\label{comparison_result}

\begin{table}[H]
  \centering
  \begin{tabular}{p{2cm}p{2.5cm}p{2.5cm}p{2.5cm}p{2.5cm}p{2.5cm}}
    \hline
    \textbf{Language} & \textbf{Agent}  & \textbf{Build} & \textbf{Management} & \textbf{Validation}  & \textbf{Manage\&Val}  \\
    \hline
    \rowcolor{gray!20}
    Python & \textbf{RepoLaunch} & \textbf{78.6\%}\cnt{33/42} & 93.9\%\cnt{31/33} & 87.1\%\cnt{27/31} & \textbf{81.8\%}\cnt{27/33} \\
    Python & SWE-agent & 52.4\%\cnt{22/42} & -- & -- & 63.6\%\cnt{14/22} \\
    \rowcolor{gray!20}
    C/C++ & \textbf{RepoLaunch} & \textbf{81.3\%}\cnt{13/16} & 100\%\cnt{13/13} & 92.3\%\cnt{12/13} & \textbf{92.3\%}\cnt{12/13} \\
    C/C++ & SWE-agent & 62.5\%\cnt{10/16} & -- & -- & 30.0\%\cnt{3/10} \\
    \rowcolor{gray!20}
    Java & \textbf{RepoLaunch} & \textbf{73.3\%}\cnt{11/15} & 81.8\%\cnt{9/11} & 88.9\%\cnt{8/9} & \textbf{72.7\%}\cnt{8/11} \\
    Java & SWE-agent & 66.7\%\cnt{10/15} & -- & -- & 40.0\%\cnt{4/10} \\
    \rowcolor{gray!20}
    JS/TS & \textbf{RepoLaunch} & \textbf{84.4\%}\cnt{27/32} & 85.2\%\cnt{23/27} & 91.3\%\cnt{21/23} & \textbf{77.8\%}\cnt{21/27} \\
    JS/TS & SWE-agent & 68.8\%\cnt{22/32} & -- & -- & 59.1\%\cnt{13/22} \\
    \rowcolor{gray!20}
    Rust & \textbf{RepoLaunch} & \textbf{70.0\%}\cnt{14/20} & 85.7\%\cnt{12/14} & 91.7\%\cnt{11/12} & \textbf{78.6\%}\cnt{11/14} \\
    Rust & SWE-agent & 65.0\%\cnt{13/20} & -- & -- & 76.9\%\cnt{10/13} \\
    \rowcolor{gray!20}
    Go & \textbf{RepoLaunch} & \textbf{81.5\%}\cnt{22/27} & 86.4\%\cnt{19/22} & 100\%\cnt{19/19} & \textbf{86.4\%}\cnt{19/22} \\
    Go & SWE-agent & 66.7\%\cnt{18/27} & -- & -- & 22.2\%\cnt{4/18} \\
    \rowcolor{gray!20}
    Overall & \textbf{RepoLaunch} & \textbf{78.9\%}\cnt{120/152} & 89.2\%\cnt{107/120} & 91.6\%\cnt{98/107} & \textbf{81.7\%}\cnt{98/120} \\
    Overall & SWE-agent & 62.5\%\cnt{95/152} & -- & -- & 50.5\%\cnt{48/95} \\
    \hline
  \end{tabular}
  \caption{
  \label{performance_comp_table}
  Performance comparison of RepoLaunch against the baseline SWE-agent, a general-purpose SWE agent equipped with only bash and file-editor tools, on the repo build and management task to reconstruct task environments and extract test statuses for SWE-bench instances \citep{jimenez2023swe,yang2025swesmith,deng2025swepro}. Here \textbf{Manage\&Val} means \textbf{Management\&Validation} combined success rate from the Build result. RepoLaunch effectively surpasses the baseline SWE-agent on both the Build and Management stages.}. 
  
\end{table}

\newpage

\begin{table*}
  \centering
  \begin{tabular}{p{2.25cm}p{1.85cm}p{2.15cm}p{2.15cm}p{2.15cm}p{2.15cm}p{2.15cm}p{2.15cm}}
    \hline
    Method&Type&Python&C/C++&Java&JS/TS&Rust&Go\\
    \hline
    \textbf{RepoLaunch}&\textbf{agentic}&
    \textbf{78.6}\%\cnt{33/42}&
    \textbf{81.3}\%\cnt{13/16}&
    \textbf{73.3}\%\cnt{11/15}&
    \textbf{84.4}\%\cnt{27/32}&
    \textbf{70.0}\%\cnt{14/20}&
    \textbf{81.5}\%\cnt{22/27} \\

    repo2run&agentic&
    59.5\%\cnt{25/42}&--&--&--&--&--\\

    CXXCrafter&agentic&
    --&68.8\%\cnt{11/16}&--&--&--&--\\

    SWE-agent&agentic&
    52.4\%\cnt{22/42}&
    62.5\%\cnt{10/16}&
    66.7\%\cnt{10/15}&
    68.8\%\cnt{22/32}&
    65.0\%\cnt{13/20}&
    66.7\%\cnt{18/27}\\

    pipreq&rule-based&
    40.5\%\cnt{17/42}&--&--&--&--\\

    railpack&rule-based&
    23.8\%\cnt{10/42}&
    31.3\%\cnt{5/16}&
    20.0\%\cnt{3/15}&
    34.4\%\cnt{11/32}&
    35.0\%\cnt{7/20}&
    44.4\%\cnt{12/27}\\
    \hline
  \end{tabular}
  \caption{Performance of different repo build methods on building repo-commit pairs of SWE-bench instances \citep{jimenez2023swe,yang2025swesmith,deng2025swepro}. repo2run, CXXCrafter, pipreq and railpack do not have a test case extractor, so only the Build stage is evaluated and the Management stage performance cannot be compared. RepoLaunch achieves state-of-the-art performance on all languages.}
  \label{performance_comp_table_build}
\end{table*}

\begin{table*}
  \centering
  \begin{tabular}{p{3cm}p{4cm}p{2.5cm}p{4cm}}
    \hline
    Method&Statistic Source&Cost in USD&Build Success Rate\\
    \hline
    \textbf{RepoLaunch}&from our experiment&\textbf{0.65}&\textbf{79\%}\\
    repo2run&from our experiment&0.70&60\%\\
    swe-rebench&\citep{badertdinov2025swe,badertdinov2026swe}&0.10&20\textasciitilde30\%\\
    OpenSWE&\citep{fu2026davinci}&19.1&\textasciitilde50\%\\
    \hline
  \end{tabular}
  \caption{Average LM API cost in USD per instance to build Python repos. RepoLaunch is cost-effective to build Python repos while achieving the highest success rate.}
  \label{cost_comp_table}
\end{table*}

\section{Ablation Study on RepoLaunch Modules and Backbone LMs}
Table~\ref{ablation} evaluates the contribution of each individual module of RepoLaunch, while Table~\ref{language_model_robustness} evaluates RepoLaunch's performance with different LM backbones. RepoLaunch remains robust with different LM backbones.

\begin{table}[H]
\centering

\small
\setlength{\tabcolsep}{4pt}
\renewcommand{\arraystretch}{1.15}

\begin{tabularx}{\textwidth}{
    >{\raggedright\arraybackslash}X
    c c c c c
}
\toprule
\textbf{Experiment}
& \makecell{\textbf{Build}\\\textbf{Success}}
& \makecell{\textbf{Management}\\\textbf{Success}}
& \makecell{\textbf{Verification}\\\textbf{Success}}
& \makecell{\textbf{Avg. Build}\\\textbf{Steps}}
& \makecell{\textbf{Avg. Manage.}\\\textbf{Steps}} \\
\midrule

\multicolumn{6}{l}{\textbf{Linux}} \\
search + prompt + image + str-replace-editor \#
& 84.9\% \cnt{129/152}
& 89.9\% \cnt{116/129}
& 92.2\% \cnt{107/116}
& 23.8
& 21.1 \\

Baseline (search + prompt + image)
& 78.9\% \cnt{120/152}
& 89.2\% \cnt{107/120}
& 91.6\% \cnt{98/107}
& 25.0
& 21.8 \\

w/o web search
& 71.7\% \cnt{109/152}
& 85.3\% \cnt{93/109}
& 91.4\% \cnt{85/93}
& 27.7
& 25.1 \\

w/o language-specific prompt
& 72.3\% \cnt{86/152}
& 80.2\% \cnt{69/86}
& 84.1\% \cnt{58/69}
& 26.1
& 24.7 \\

w/o language-specific image$^\ast$
& 67.1\% \cnt{102/152}
& 88.2\% \cnt{90/102}
& 91.1\% \cnt{82/90}
& 31.1
& 22.5 \\

\midrule

\multicolumn{6}{l}{\textbf{Windows}} \\
search + prompt + image + str-replace-editor \#
& 68.0\% \cnt{136/200}
& 72.8\% \cnt{99/136}
& 80.8\% \cnt{80/99}
& 36.7
& 29.8 \\

Baseline (search + prompt + image)
& 60.5\% \cnt{121/200}
& 70.2\% \cnt{85/121}
& 80.0\% \cnt{68/85}
& 41.2
& 31.7 \\

w/o web search
& 51.5\% \cnt{103/200}
& 66.0\% \cnt{68/103}
& 80.9\% \cnt{55/68}
& 48.6
& 39.8 \\

w/o language-specific prompt
& 52.5\% \cnt{105/200}
& 61.9\% \cnt{65/105}
& 73.8\% \cnt{48/65}
& 47.6
& 38.2 \\

w/o language-specific image$^\ast$
& 46.0\% \cnt{92/200}
& 70.7\% \cnt{65/92}
& 80.0\% \cnt{52/65}
& 58.6
& 32.9 \\

\bottomrule
\end{tabularx}

\caption{Ablation results on Linux and Windows. The baseline is the setting adopted in other experiments in the paper. \# The string-replace-editor tool is the tool to edit file contents in the repos. As RepoLaunch is used to build environments for SWE tasks at base commits with issues, by default the file editor tool is disabled to prevent the agent from fixing the bugs originally in the repos. * When language-specific image is disabled, RepoLaunch builds all repos from a raw Ubuntu (for Linux) or Windows-Server (for Windows) base image with no other dependencies installed.}
\label{ablation}
\end{table}

\begin{table}[H]
  \centering
  \begin{tabular}{p{3.0cm}p{2.5cm}p{2.5cm}p{2.5cm}p{2.5cm}p{2.5cm}}
    \hline
    Model&Agent&Build success&Management success&Validation Success&Manage\&Val success\\
    \hline
    \rowcolor{gray!20}
    Claude-Sonnet-5&\textbf{RepoLaunch}&\textbf{90.8}\%\cnt{138/152}&95.7\%\cnt{132/138}&96.2\%\cnt{127/132}&\textbf{92.0}\%\cnt{127/138}\\
    Claude-Sonnet-5&SWE-agent&81.6\%\cnt{124/152}&-&-&64.5\%\cnt{80/124}\\
    Claude-Sonnet-5&repo2run*&69.0\%\cnt{29/42}&-&-&-\\
    \rowcolor{gray!20}
    Deepseek-v4-Pro&\textbf{RepoLaunch}&\textbf{84.9}\%\cnt{129/152}&92.2\%\cnt{119/129}&89.1\%\cnt{106/119}&\textbf{82.1}\%\cnt{106/129}\\
    Deepseek-v4-Pro&SWE-agent&77.6\%\cnt{118/152}&-&-&63.6\%\cnt{75/118}\\
    Deepseek-v4-Pro&repo2run&71.4\%\cnt{30/42}&-&-&-\\
    \rowcolor{gray!20} 
    GPT-5-Medium&\textbf{RepoLaunch}&\textbf{78.9}\%\cnt{120/152}&89.2\%\cnt{107/120}&91.6\%\cnt{98/107}&\textbf{81.7}\%\cnt{98/120}\\
    GPT-5-Medium&SWE-agent&62.5\%\cnt{95/152}&-&-&50.5\%\cnt{48/95}\\
    GPT-5-Medium&repo2run&59.5\%\cnt{25/42}&-&-&-\\
    \hline
  \end{tabular}
  \caption{RepoLaunch is robust when driven by different LM providers. RepoLaunch and SWE-agent are evaluated on the whole multilingual benchmark in Appendix~\ref{comparison_instance_list} on both Build and Manage\&Val stages. * repo2run can only run on the Python subset of the benchmark on the Build stage.}
  \label{language_model_robustness}
\end{table}

\section{Cross-platform Transferability of RepoLaunch framework}

\label{transferability}

\subsection{Result}

The RepoLaunch framework is designed to adapt to any language and any platform. To adapt it to a new language, one simply needs to provide language-specific prompts for build and test; however, as recent LMs possess greater internal knowledge of a language, this prompt is becoming less useful. To adapt it to a new platform, only the base operating system image for that platform is required. As most platforms other than Windows are Unix-based, their base images can be built from Linux images, making cross-platform transferability easy.

To demonstrate the RepoLaunch framework’s cross-platform compatibility, we have also tested it on Android and embedded systems. Although we intend to test it on macOS/iOS, we found that no public macOS/iOS image with a CLI interface is currently available, mainly because Apple’s policy has not permitted the publication of macOS/iOS development containers so far. We were therefore unable to carry out tests on macOS / iOS.

\textbf{Android Result.} We sampled 25 repos for Android mobile systems. The repo list is in Appendix \ref{android_repo_list}. We adopt publicly-available Android system base images from provider \textit{cimg/android}, which are built from Linux base images. Experiments are done with GPT-5-Thinking-Medium.

Build stage success 23/25 = 92\%

Management stage success 23/23 = 100\% 

Overall success 23/25 = 92\%

\textbf{Embedded system result. }We examined popular public repositories for embedded systems, such as those for Raspi. We found that most of these repositories are still Linux-based, with a small minority based on Windows. However, Windows-based embedded repositories are often distributed as pre-built executables without regression tests (e.g., microsoft/Windows-IoT-Samples, microsoft/Windows-iotcore-samples, and ms-iot/windows-iotent-deploys), making them unsuitable for our evaluation setting. As a result, we sampled 25 Linux-based embedded repositories for evaluation. The full repo list is in Appendix \ref{embed_repo_list}. The evaluation still uses Linux based images. RepoLaunch installs specific dependencies of embedded applications during build.

Build stage success 22/25 = 88\%

Management stage success 22/22 = 100\% 

Overall success 22/25 = 88\%

\subsection{Android repo test set.}
\label{android_repo_list}

Please refer to the left column in the below table.

\subsection{Embedded system repo test set.}
\label{embed_repo_list}

Please refer to the right column in the below table.

\setlength{\tabcolsep}{0.1mm}
\small
\begin{longtable}{p{0.5cm}p{5cm}p{3cm}p{0.5cm}p{5.1cm}p{1.5cm}}
\toprule
\textbf{id} & \textbf{repo} & \textbf{language} & \textbf{id} & \textbf{repo} & \textbf{language} \\ 
\midrule
\endfirsthead

\multicolumn{6}{c}%
{{continued from previous page}} \\
\toprule
\textbf{id} & \textbf{repo} & \textbf{language} & \textbf{id} & \textbf{repo} & \textbf{language} \\
\midrule
\endhead

\midrule \multicolumn{6}{r}{{Continued on next page}} \\ \midrule
\endfoot

\bottomrule
\endlastfoot
\multicolumn{3}{c}{\textbf{Android repos}} & \multicolumn{3}{c}{\textbf{Embedded system repos}} \\
1&robolectric/robolectric & Java & 1&raspberrypi/linux&C\\
2&skylot/jadx & Java & 2&Koenkk/zigbee2mqtt&TypeScript\\
3&JesusFreke/smali & Java & 3&esphome/esphome&C++\\
4&zxing/zxing & Java &4&home-assistant/core&Python\\
5&google/bundletool & Java & 5&node-red/node-red&JavaScript\\
6&iBotPeaches/Apktool & Java & 6&Arksine/moonraker&Python\\
7&pxb1988/dex2jar & Java & 7&homebridge/homebridge&TypeScript\\
8&apache/cordova-android & JS + Java & 8&blakeblackshear/frigate&TypeScript\\
9&appium/appium-uiautomator2-driver & TS & 9 & AdguardTeam/AdGuardHome&Go\\
10&appium/appium & JS+TS & 10&OctoPrint/OctoPrint&Python\\
11&square/retrofit & Java & 11&motioneye-project/motioneye&Python\\
12&square/okhttp & Java + Kotlin & 12&Klipper3d/klipper&C\\
13&ReactiveX/RxAndroid & Java & 13&mendersoftware/mender&C++\\
14&bumptech/glide & Java & 14&mopidy/mopidy&Python\\
15&greenrobot/EventBus & Java & 15&openwrt/openwrt&C\\
16&Konloch/bytecode-viewer & Java & 16&gpiozero/gpiozero&Python\\
17&TeamNewPipe/NewPipeExtractor & Java & 17&domoticz/domoticz&C++\\
18&wix/Detox & JS + TS + Java & 18&adafruit/Adafruit\_Blinka&Python\\
19&ionic-team/capacitor & TS + Java & 19&brgl/libgpiod&C++\\
20&android/testing-samples & Java & 20&eclipse-kura/kura&Java\\
21&kivy/buildozer & Python & 21&k3s-io/k3s&Go\\
22&kivy/python-for-android & C + Python & 22&sbabic/swupdate&C\\
23&beeware/briefcase & Python & 23&Pi4J/pi4j&Java\\
24&facebook/redex & C++ + Python & 24&vsergeev/c-periphery&C \\
25&tauri-apps/tauri & Rust + TS + Kotlin & 25&vsergeev/python-periphery&Python

\end{longtable}

\section{Benchmarking Result of SWE-bench-Live/MultiLang and Windows}

Please refer to Table \ref{eval_res} and \ref{windows_benchmarking_table}.

\begin{table}[H]
  \centering
  \begin{tabular}{p{2.2cm}p{3.5cm}p{1.3cm}p{1.3cm}p{1.3cm}p{1.3cm}p{1.3cm}p{1.3cm}p{1.3cm}p{1.3cm}}
    \hline
    \textbf{Agent}  & \textbf{Model} & \textbf{C/C++}  & \textbf{C\#} & \textbf{Java} & \textbf{JS/TS} & \textbf{Go} & \textbf{Rust} & \textbf{Linux}  \\
    \hline
    SWE-agent & GPT-5.5 (Medium) & 34.8\% & \textbf{44.9\%} & \textbf{50.0\%} & \textbf{47.5\%} & 34.7\% & 28.1\% & \textbf{40.0\%} \\
    SWE-agent & Deepseek-v4-Pro & 24.2\% & \textbf{44.9}\% & 39.2\% & 36.8\% & 29.6\% & 34.4\% & 34.8\% \\
    SWE-agent & Claude-4.5-Sonnet & \textbf{43.8}\% & 20.5\% & 16.1\% & 20.4\% & 39.7\% & 28.9\% & 28.4\% \\
    SWE-agent & GPT-5.2 (Medium) & 41.7\% & 21.8\% & 22.6\% & 16.1\% & 25.0\% & 28.9\% & 26.0\% \\
    SWE-agent & Gemini-3-flash & 35.4\% & 14.3\% & 25.8\% & 9.9\% & 32.4\% & \textbf{37.5\%} & 25.9\%  \\
    SWE-agent & Deepseek-v3.1-terminus & 27.1\% & 21.8\% & 27.4\% & 10.5\% & 20.6\% & 28.9\% & 22.7\%  \\
    OpenHands & GPT-5.5 (Medium) & 33.3\% & 42.3\% & 47.1\% & 45.1\% & 25.5\% & 29.7\% & 37.2\%  \\
    OpenHands & Deepseek-v4-Pro & 22.7\% & 38.5\% & 39.2\% & 36.8\% & 27.6\% & 28.1\% & 32.2\%  \\
    OpenHands & Claude-4.5-Sonnet & \textbf{43.8}\% & 17.9\% & 19.4\% & 19.1\% & 39.7\% & 26.7\% & 27.8\%  \\
    OpenHands & GPT-5.2 (Medium) & 37.5\% & 20.5\% & 22.6\% & 13.0\% & 20.6\% & 26.7\% & 23.6\%  \\
    OpenHands & Gemini-3-flash & 35.4\% & 17.9\% & 24.2\% & 8.6\% & 30.9\% & 35.6\% & 25.4\%  \\
    OpenHands & Deepseek-v3.1-terminus & 25.0\% & 17.9\% & 24.2\% & 11.1\% & 22.1\% & 26.7\% & 21.2\%  \\
    ClaudeCode & GPT-5.5 (Medium) & 37.9\% & 39.7\% & 45.1\% & 42.6\% & 37.8\% & 32.8\% & 39.3\%  \\
    ClaudeCode & Deepseek-v4-Pro & 22.7\% & 37.2\% & 38.2\% & 30.9\% & 29.6\% & 29.7\% & 31.4\%  \\
    ClaudeCode & Claude-4.5-Sonnet & 35.4\% & 14.3\% & 27.4\% & 27.2\% & \textbf{43.8\%} & 22.2\% & 28.4\%  \\
    ClaudeCode & GPT-5.2 (Medium) & 37.5\% & 14.3\% & 37.1\% & 20.4\% & 27.9\% & 33.3\% & 28.4\%  \\
    ClaudeCode & Deepseek-v3.1-terminus & 10.4\% & 14.3\% & 22.6\% & 15.4\% & 29.4\% & 17.8\% & 18.3\%  \\

    \hline
  \end{tabular}
  \caption{\label{eval_res}
    The success rate of different model-agent combinations on SWE-bench-Live/MultiLang. The Linux column is the average values of previous columns so that languages with different number of instances have the same weight.
  }
\end{table}

\begin{table}[H]
  \centering
  \begin{tabular}{p{3cm}p{2cm}p{2cm}p{2cm}p{2cm}p{2cm}p{2cm}}
    \hline
    \textbf{LM} & Claude-4.5 & GPT-5.5 & GPT-5.2 & Gemini-3 & Ds-4 &  Ds-3.1  \\
    \hline
    \textbf{Agent} & Win-agent* & Win-agent & Win-agent &  Win-agent & Win-agent &  Win-agent \\
    \hline
    \textbf{Resolved (\%)} & \textbf{30.0\%} & 26.7\% & 20.0\% & 16.7\% & 21.7\% &  20.0\% \\
    \hline
  \end{tabular}
  \caption{\label{windows_benchmarking_table}
    Benchmarking results of latest LMs on SWE-bench-Live/Windows. GPT models use medium reasoning effort. Claude-4.5: Claude-4.5-Sonnet, Gemini-3: Gemini-3-Flash, Ds-3.1: DeepSeek-v3.1-Terminus, Ds-4: Deepseek-v4-Pro. Win-agent* is the minimal agent implemented by our team with the same tool calls as SWE-agent and OpenHands. It is also the first coding agent that can run on Windows containers.
  }
\end{table}

\newpage

\begin{figure*}[!t]
  \includegraphics[width=1.0\linewidth]{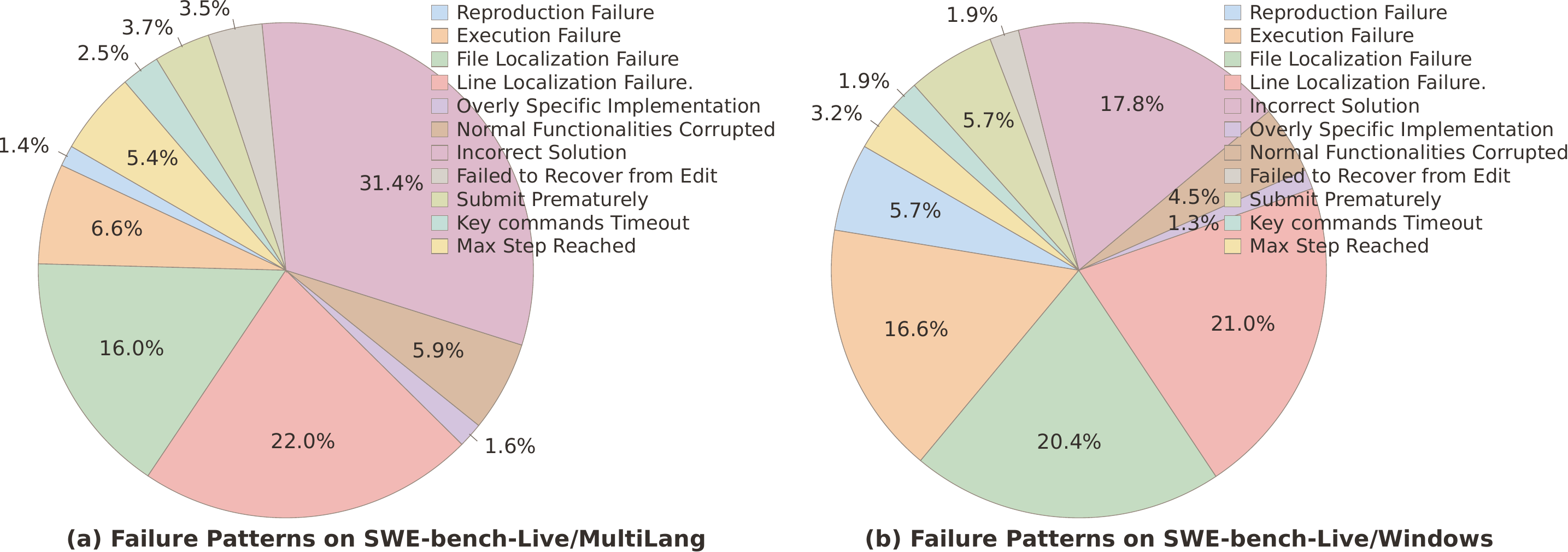}
  \caption {The proportion of different failure patterns of coding agents on SWE-bench-Live.}
  \label{failue_pattern_swe_bench_live}
\end{figure*}

\section{Failure Pattern Analysis of Coding Agents' Trajectories on SWE-bench-Live/MultiLang \& Windows}
\label{pattern_solution}

Following the failure pattern analysis of SWE-agent \citep{yang2024swe} and ORACLE-SWE \citep{li2026oracle}, and through further case study of 100 failed trajectories of the coding agents evaluated on SWE-bench-Live/MultiLang and Windows, we propose the following failure patterns when solving SWE-bench-Live:

\begin{enumerate}[nosep]
\item \textbf{Reproduction Failure.} The reproduction test file written by agent runs successfully but cannot reproduce the issue.
\item \textbf{Execution Failure.} The agent even does not know how to run the code repository, so it fails to run the written reproduction test file.
\item \textbf{File Localization Failure.} The agent did not view all the files that need to be edited in ground truth patch.
\item \textbf{Line Localization Failure.} The agent failed to edit correct locations that the ground truth patch edits.
\item \textbf{Overly Specific Implementation}.
\item \textbf{Normal Functionality Corrupted.} The agent's edits corrupted originally normal functionalities.
\item \textbf{Incorrect Solution} to the problem statement.
\item \textbf{Failed to Recover from Edit.} The agent fell into edit-error-edit-error loop.
\item \textbf{Submit Prematurely}.
\item \textbf{Key commands Timeout.} Running important shell commands like testing or modifying dependencies went timeout so verification of edits is not possible.
\item \textbf{Max Step Reached}, before completing the task.
\end{enumerate}

We then use GPT-5.2-Thinking-Medium to classify all the failed trajectories on SWE-bench-Live/MultiLang and Windows. Result is shown in Figure \ref{failue_pattern_swe_bench_live}. The discrepancy between manual labeling and LM labeling does not exceed 20\%. Failure pattern on Linux illustrates that:

\begin{enumerate}[label=(\arabic*), nosep]
\item The reproduction of the problem and the validation of the fix mostly fails to work because the agents do not know how to run the code repo.
\item Failure of localization is a major reason of failure.
\item Normal functionality corrupted by agents' fix also accounts for 5.90\%, showing it is also important to instruct agents to find the correct way to run regression tests in the repo.
\item Only 5.41\% percent of the instances are tagged as max step reached, showing max step of 100 is mostly enough.
\item The known problem of LM agents falling into edit-failure loop seems to be mitigated by latest stronger LMs, with the proportion dropped from 23.4\% in the original SWE-agent paper \citep{yang2024swe} for GPT-4 to less than 4\% for models evaluated in this paper. 
\end{enumerate}

Failure pattern on Windows illustrates that:

\begin{enumerate}[label=(\arabic*), nosep]
\item The pattern mostly aligns with that on Linux, with localization as the major problem and very few instances reaching max step limit.
\item The proportions of failures to reproduce the issue and execute code both increased significantly, showing that Windows tasks are more difficult in terms of configuring the settings, managing the executable environment and running code successfully.
\item However, the proportion of implementation problems (such as being overly specific, original functionality corrupted, or fix incorrectly implemented) drops significantly.
From our initial observation, these Windows tasks are relatively easy in code implementation -- mostly writing / fixing the corresponding Windows-specific code given the existing sound implementation in Linux, such as modifying file path from Unix format to Windows format and modifying commands from bash commands to powershell commands.
The trivial nature of this aspect makes the observed drop in success rate relatively modest.
As for future work, we plan to conduct a more comprehensive analysis of the categories and recurring patterns of Windows tasks derived from GitHub issues.
\end{enumerate}

\end{document}